# A Self-Replicating Peptide Nucleic Acid

by **Tobias A. Plöger** and **Günter von Kiedrowski**

Lehrstuhl für Organische Chemie I, Bioorganische Chemie, Ruhr-Universität Bochum,
Universitätsstraße 150, NC 2/173, D-44780 Bochum
(phone: +49(0)234 32-28218; fax: +49(0)234 32-14355; tobias@oc1.rub.de)

We report on a case of autocatalytic feedback in a template directed synthesis of a self-complementary hexa-PNA from two trimeric building blocks. The course of the reaction was monitored in the presence of increasing initial concentrations of product by means of RP-HPLC. Kinetic modeling with the *SimFit* program revealed parabolic growth according to the so-called square-root law. The observed template effect, as well as the rate of the ligation, was significantly influenced by factors like nucleophilic catalysts, pH value, and uncharged co-solvents. Systematic optimization of the reaction conditions allowed us to increase the autocatalytic efficiency of the system by two orders of magnitude.

**1. Introduction.** – Today's DNA/RNA/protein-world exhibits a degree of complexity that precludes a *de novo* synthesis under prebiotic conditions. More conceivable seems the emergence of an evolvable system that provides genetic storage as well as catalytic activity within the same molecule. Supported by the discovery of ribozymes [1-4] and reverse transcriptases [5, 6], it is now widely believed that a world solely based on RNA preceded our contemporary life [7-11]. This RNA-world scenario has, however, problems of its own [12, 13]: (i) the prebiotic synthesis of sufficient amounts of ribose and ribonucleosides, (ii) the long-term chemical stability of RNA oligomers, and (iii) the origin of long homochiral oligomers due to enantiomeric cross-inhibition during template-directed polymerizations.

Therefore, it was suggested that the RNA-world was preceded by one or more simpler and thus chemically more stable genetic material [13-16]. One appealing candidate for this is PNA, a DNA or RNA mimic based on a non-charged, achiral, and pseudopeptidic backbone consisting of *N*-(2-aminoethyl)glycine units [17-25]: Building blocks for PNA and related peptide nucleic acids were identified in prebiotic model experiments [26, 27] and amongst the organic components of the *Murchison* meteorite [28]. Further, the achiral backbone offers a path for a gradually transformation to a homochiral polymer while avoiding enantiomeric cross-inhibition [20, 21, 29]. Finally, genetic takeover from PNA to RNA and from DNA to PNA has been demonstrated as well as templated PNA-ligation and the formation of DNA-PNA-chimeras on PNA- or DNA-templates, respectively [30-40]. On the other hand, cases for PNA (self-)replication as a prerequisite for natural selection and evolution have not been identified so far. Recently, we designed a system to evaluate potential PNA self-replication and presented an efficient large-scale synthesis of the building blocks needed [41]. This system consists of two trimeric building blocks F-PNA **A** and PNA **B** leading to the self-complementary hexa-PNA **T** upon ligation (*Scheme 1*); consequently it will be able to compare this system with earlier studies from our laboratory [42-46]. Natural thymine was replaced by 2,4-difluorotoluene to introduce a $^{19}$F-NMR probe to study the system by kinetic NMR titration [47] in the long term. Furthermore, the C- and N-termini were modified with novel unreactive solubility enhancers which are positively charged at physiological pH and avoid the introduction of stereogenic centers.

Scheme 1. *Self-Replicating System on the Basis of PNA*. Trimeric building blocks **A** and **B** give the self-complementary hexa-PNA **T** upon condensation. Green: solubility enhancer; red: ligation site; yellow: fluorine label.

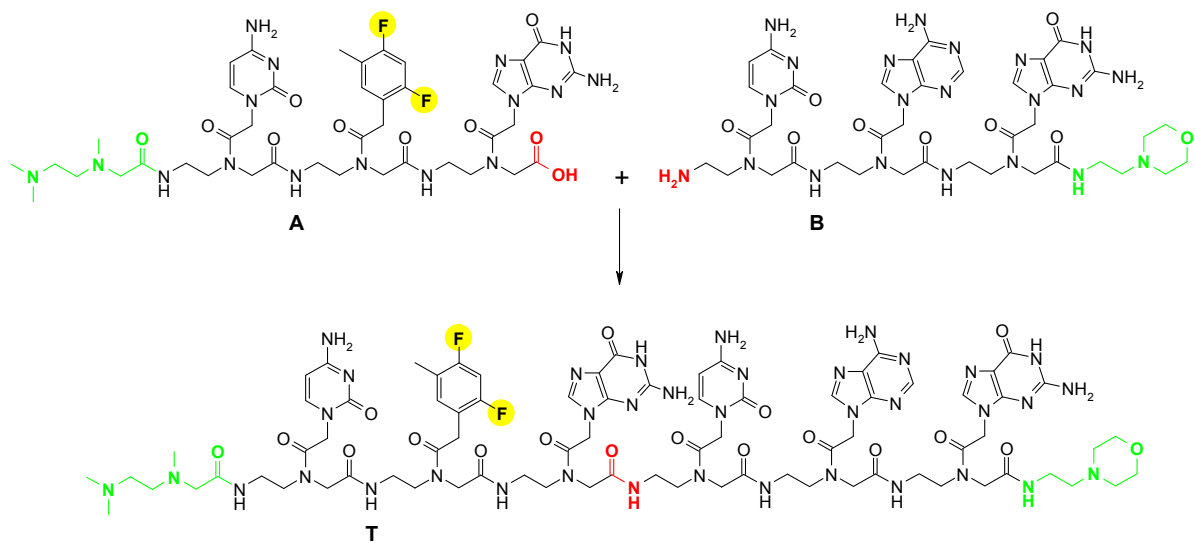

**2. Design of the Study.** – The ligation of building blocks F-PNA **A** and PNA **B** in aqueous buffered solutions was monitored by RP-HPLC. In order to detect self-replication, the reaction was followed in the presence of different amounts of initially added ligation product **T** [48]. The resulting data points were then simultaneously kinetically modeled using our *SimFit* program [49]: Model A (*eqs. 1* and *2*) includes the bimolecular background reaction of **A** and **B** to give hexa-PNA **T** ($k_a$, *Scheme 2*) as well as a (pseudo) first order decay of F-PNA **A** ($k_c$) which allows to correct for the hydrolysis of the coupling reagent.

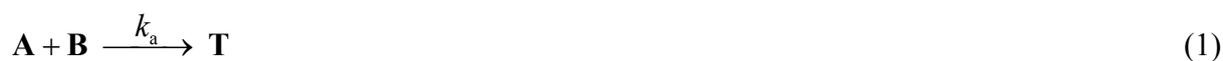

$$\mathbf{A} + \mathbf{B} \xrightarrow{k_a} \mathbf{T} \tag{1}$$

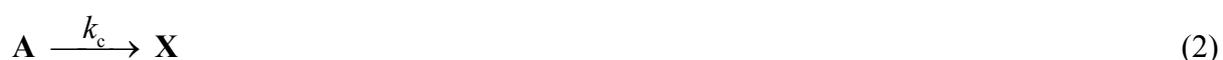

$$\mathbf{A} \xrightarrow{k_c} \mathbf{X} \tag{2}$$

Model B (*eqs. 1-3*) considers also an autocatalytic pathway in which **T** is supposed to be able to reversibly preorganize a pair of **A** and **B** ($K_1$, *Scheme 2*) by hydrogen bonding in a termolecular complex [**A·B·T**]. In consequence, a pseudo-unimolecular ligation reaction ($k_b$) should take place leading to template duplex [**T·T**]. The reversible dissociation ($K_2$) of this duplex produces two template molecules **T**, which may both enter another replication cycle. However, from thermodynamic considerations and data for published replication systems, it can be concluded, that most template molecules **T** will be present as [**T·T**], a small fraction as [**A·B·T**], and another small portion in an unpaired state. Due to this so-called product inhibition, one can expect the system to reveal parabolic growth (factor 0.5, *eq. 3*). Further information on this topic can be found in [48].

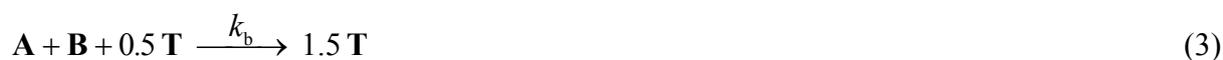

$$\mathbf{A} + \mathbf{B} + 0.5\,\mathbf{T} \xrightarrow{k_b} 1.5\,\mathbf{T} \tag{3}$$

A better agreement of the experimental data with model B than with A, as defined by the root mean square (RMS), was consequently taken as an indication for self-replication. Its efficiency under different reaction conditions was then determined by the autocatalytic excess factor $\varepsilon$ (*eq. 4*), the quotient of the rate constants of the autocatalytic reaction channel and the background reaction.

$$\varepsilon = \frac{k_b}{k_a} \tag{4}$$

In addition, the covariance of the variable rate constants gave information about the reliability of the results, with an absolute value below 0.9 indicating an acceptable independency. All figures presented show theoretical curves based on model B to ensure comparability of the results.

Scheme 2. *A Minimal Model of a Self-Replicating System*

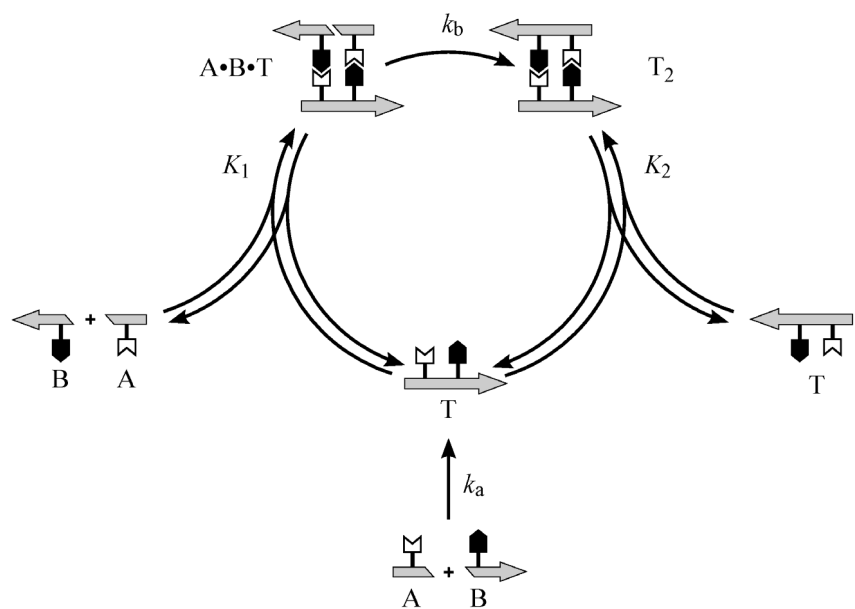

**3. Results and Discussion.** – 3.1 *Ligation and Replication in Imidazole Buffers*. Our initial assays used 5 mM building blocks **A** and **B**, 200 equivalents of the water-soluble carbodiimide EDC (0.2 M), and 0.4 M imidazole/HCl buffers at pH 7.5. These conditions resemble those used in some of the studies on templated PNA-ligation [30, 31]. The given imidazole derivative does not only regulate the pH, but also influences the ligation by acting as a nucleophilic catalyst (*Scheme 3*): First, an *O*-acylisourea **A*** is generated from the reaction of the carboxylic acid **A** with EDC. This active species can then react with the imidazole **1** to give the more stable imidazolide **A**** which yields the condensation product **T** upon reaction with the amine **B**. Consequently, the choice of the imidazole buffer is a crucial factor: The first report on PNA ligation [30] revealed EDC mediated coupling to be efficient in imidazole buffer, while 2-methylimidazole buffer inhibited the reaction. Nevertheless, EDC/2-methylimidazole was used in a further study by the same groups [31]. The influence of different imidazole derivatives on the efficiency of template-directed reactions is already known from the work of *Orgel* on RNA [50], but there has been no systematical search through a library of imidazole derivatives for PNA, and therefore studies on PNA ligation predominantly employ the EDC/imidazole system [25, 35, 36].

Scheme 3. *Mechanism Purposed for EDC Mediated and Imidazole Catalyzed PNA Ligation*

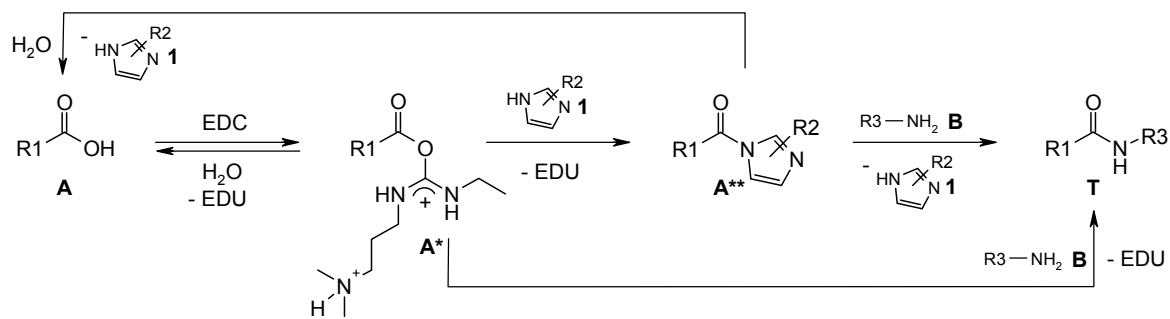

In the presence of 1-methylimidazole buffer no template effect could be observed at 0° (*Fig. 1a*); accordingly model A fits the data better than model B and $\varepsilon$ has a value close to zero (*Table 1*). Contrary, when imidazole buffer is used, one can observe a positive influence of added template on rate and yield of the reaction (*Fig. 1b*) which is also reflected in the kinetic data ($\varepsilon = 12$ M$^{-1/2}$, *Table 1*).

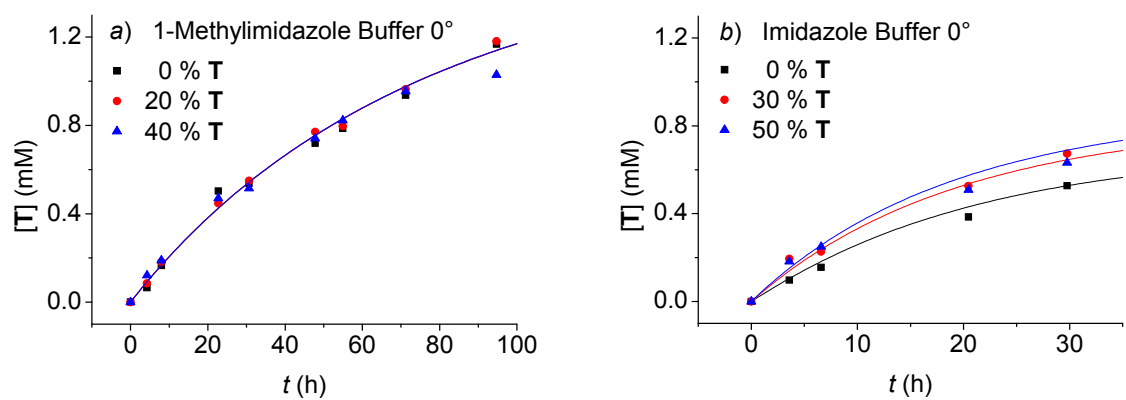

Fig. 1. *Kinetics of PNA Self-Replication Assays Involving Two Different Imidazole Buffers at 0°*. Shown are the experimental data points and fittings according to model B. Conditions: 5 mM tri-PNAs **A** and **B**, 0.2 M EDC, a: 0.4 M 1-methylimidazole pH 7.5, b: 0.4 M imidazole pH 7.5, 0°, 25 µl scale. a: cov($k_a$, $k_b$) = -0.626, cov($k_a$, $k_c$) = 0.593, and cov($k_b$, $k_c$) = 0.204. b: cov($k_a$, $k_b$) = -0.233, cov($k_a$, $k_c$) = 0.630, and cov($k_b$, $k_c$) = 0.568.

Table 1. *Kinetic Data for PNA Self-Replication Assays Involving Two Different Imidazole Buffers at 0°*[a])

| Buffer | Model | RMS (%) | $k_a$ (M$^{-1}$ s$^{-1}$) | $k_b$ (M$^{-3/2}$ s$^{-1}$) | $k_c$ (s$^{-1}$) | $\varepsilon$ (M$^{-1/2}$) |
|---|---|---|---|---|---|---|
| 1-MeIm | A | 1.18 | (2.60±0.05)×10$^{-4}$ | - | (2.4±0.2)×10$^{-6}$ | - |
|  | B | 1.22 | (2.47±0.07)×10$^{-4}$ | (0.2±1.4)×10$^{-4}$ | (2.0±0.2)×10$^{-6}$ | 0.08 |
| Im | A | 1.75 | (4.5±0.4)×10$^{-4}$ | - | (1.0±0.2)×10$^{-5}$ | - |
|  | B | 0.976 | (3.3±0.2)×10$^{-4}$ | (3.8±0.5)×10$^{-3}$ | (1.1±0.1)×10$^{-5}$ | 12 |

[a]) 5 mM tri-PNAs **A** and **B**, 0.2 M EDC, 0.4 M buffer pH 7.5, 0°, 25 µl scale

After raising the temperature to 10°, a small template effect can be detected in 1-methylimidazole buffer, whereas the efficiency in imidazole buffer is more than doubled (*Fig. 2*, *Table 2*). In general, every replicator exhibits a characteristic optimal temperature because both, the rate constant $k_b$, and the equilibrium concentration of the termolecular complex [**A·B·C**] depend on the temperature: The rate constant of the ligation is expected to increase with the temperature, while the equilibrium concentration of [**A·B·C**] is expected to decrease due to the weakening of the hydrogen bonds which stabilize the complex. From theoretical considerations, one can expect a parabolic replicator to work at its maximum rate if the temperature is adjusted between the melting temperatures of the termolecular complex and the product duplex [48]. The latter could not be determined exactly due to a non-cooperative melting process that presumably results from the disturbing effect of the fluoroaromatic isostere [51] (see *suppl.*). Therefore, we choose 10° for the following experiments to ensure a sufficient population of the productive termolecular complex. Interestingly, the slower assay displays the more pronounced template effect here, while in case of DNA replicators the autocatalytic efficiency was found to be enhanced by faster ligation chemistries [42-44, 52].

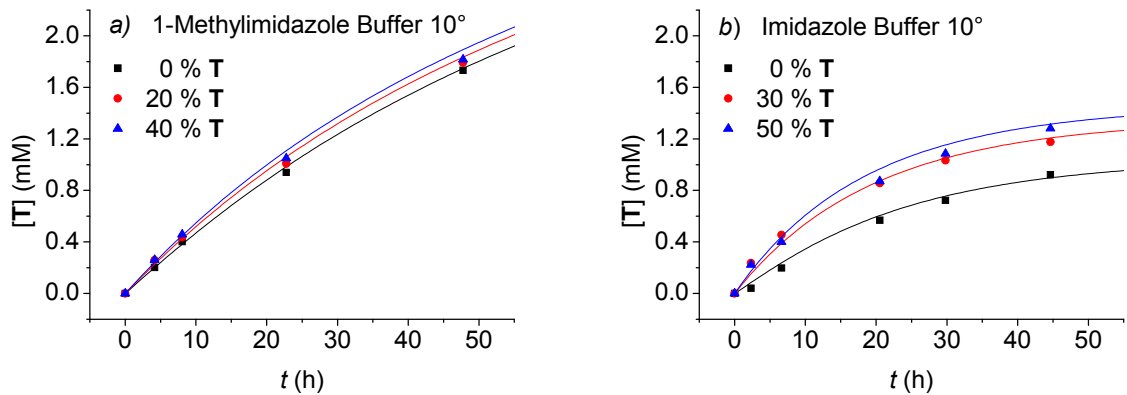

Fig 2. *Kinetics of PNA Self-Replication Assays Involving Two Different Imidazole Buffers at 10°*. Shown are the experimental data points and fittings according to model B. Conditions: 5 mM tri-PNAs **A** and **B**, 0.2 M EDC, a: 0.4 M 1-methylimidazole pH 7.5, b: 0.4 M imidazole pH 7.5, 10°, 25 μl scale. a: cov($k_a$, $k_b$) = -0.712, cov($k_a$, $k_c$) = 0.252, and cov($k_b$, $k_c$) = 0.454. b: cov($k_a$, $k_b$) = -0.202, cov($k_a$, $k_c$) = 0.616, and cov($k_b$, $k_c$) = 0.587.

Table 2. *Kinetic Data for PNA Self-Replication Assays Involving Two Different Imidazole Buffers at 10°*[a])

| Buffer | Model | RMS (%) | $k_a$ (M$^{-1}$ s$^{-1}$) | $k_b$ (M$^{-3/2}$ s$^{-1}$) | $k_c$ (s$^{-1}$) | $\varepsilon$ (M$^{-1/2}$) |
|---|---|---|---|---|---|---|
| 1-MeIm | A | 1.40 | (6.4±0.2)×10$^{-4}$ | - | (0.7±2.7)×10$^{-7}$ | - |
| | B | 0.774 | (5.3±0.1)×10$^{-4}$ | (3.3± 0.3)×10$^{-3}$ | (0.08±1.54)×10$^{-7}$ | 6.2 |
| Im | A | 4.20 | (7.2±0.7)×10$^{-4}$ | - | (8±2)×10$^{-6}$ | - |
| | B | 1.32 | (3.6±0.2)×10$^{-4}$ | (1.06±0.05)×10$^{-2}$ | (9.9±0.6)×10$^{-6}$ | 29 |

[a]) 5 mM tri-PNAs **A** and **B**, 0.2 M EDC, 0.4 M buffer pH 7.5, 10°, 25 μl scale

Noteworthy, the deviation of the theoretical curves from the experimental data increases with the amount of initially added template **T**. This stems from the hydrolysis of the coupling reagent EDC which is effectively catalyzed by carboxylic acids: While the concentration of PNA carboxylic acid **B** is kept constant at 5 mM in all experiments, the quantity of trifluoroacetic acid correlates with the initial template concentration because all PNAs were used as trifluoroacetates owing to HPLC purification. In fact, both models account for EDC hydrolysis in general (*eq. 2*) but do not consider its dependency on template concentration. In addition, following the reaction necessitates the calculation of differences between HPLC integrals which is reasoned to give higher errors in the presence of elevated amounts of initial template.

3.2 *Ligation and Replication in MOPS Buffer*. Further assays employed MOPS buffer, thus we were also able to investigate nucleophilic catalysts that cannot function as buffers in the neutral pH range. Generally, replication experiments at 10° were run simultaneously with control experiment at room temperature. In the absence of any added catalysts, efficient replication took place (*Fig. 3a, Table 3*), whereas significant amounts of side products were detected (data not shown). These could be suppressed effectively by adding any of the heterocyclic catalysts studied. Addition of imidazole or pyridine maintained self-replication at a reduced efficiency of 40 or 19 $M^{-1/2}$, respectively, while 1-methyl-imidazole and HOAt suppressed hexamer formation *via* the instructed reaction channel. The latter ones, as well as pyridine, could rather be identified as efficient catalysts for PNA-ligation at room temperature.

In contrast to DNA and RNA complexes, whose structures and stabilities are drastically governed by salt concentration and the valency of the cations [53-57], PNA duplexes are typically independent to these factors owing to the non-charged backbone [58]. However, we reasoned that different anions might have an influence on the replication system because of the positively charged end modifications (*Fig. 4, Table 4*). All experiments already contained 0.2 M chloride ions due to the use of EDC hydrochloride as condensing reagent. At 0.1 M additionally chloride, neither the efficiency nor the yield of the replication were considerably affected. In contrast, both, sulfate and iodide improved the efficiency of the replication to a similar extent. Sulfate addition decreased the yield of hexa-PNA **T**, while iodide had a positive influence.

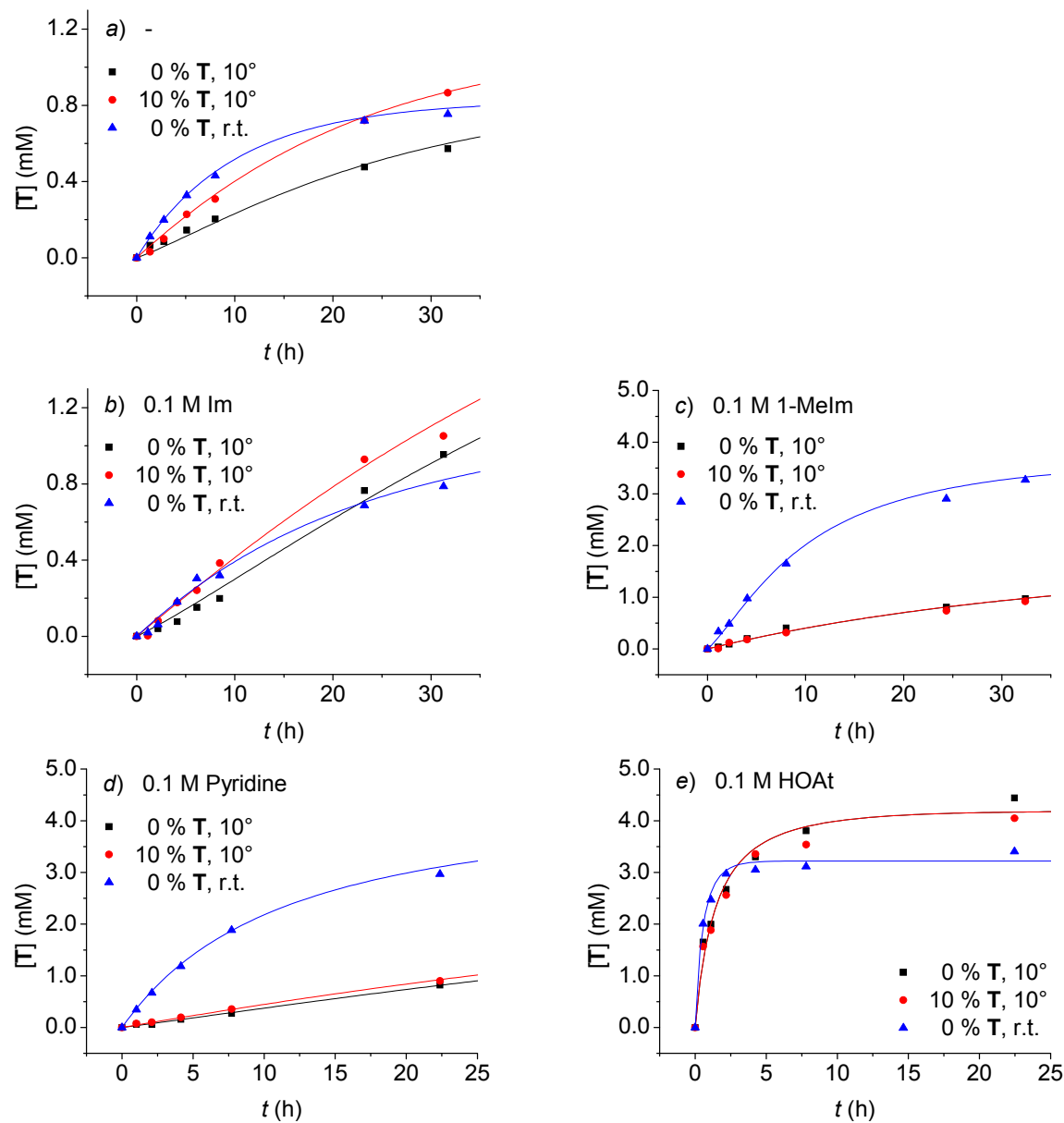

Fig 3. *Kinetics of PNA Self-Replication Assays Involving Different Nucleophilic Catalysts.* Shown are the experimental data points and fittings according to model B. Conditions: 5 mM tri-PNAs **A** and **B**, 0.2 M EDC, 0.2 M MOPS pH 7.2, 0.2 M NaCl, b: 0.1 M imidazole, c: 0.1 M 1-methylimidazole, d: 0.1 M pyridine, e: 0.1 M HOAt, 6 µl scale. a/10°: cov($k_a$, $k_b$) = -0.458, cov($k_a$, $k_c$) = 0.101, and cov($k_b$, $k_c$) = 0.815; a/r.t.: cov($k_a$, $k_b$) = -0.960, cov($k_a$, $k_c$) = -0.781, and cov($k_b$, $k_c$) = 0.920. b/10°: cov($k_a$, $k_b$) = -0.492, cov($k_a$, $k_c$) = 0.355, and cov($k_b$, $k_c$) = 0.596; b/r.t.: cov($k_a$, $k_b$) = 0.877, cov($k_a$, $k_c$) = 2.327, and cov($k_b$, $k_c$) = -0.825. c/10°: cov($k_a$, $k_b$) = -0.494, cov($k_a$, $k_c$) = 0.463, and cov($k_b$, $k_c$) = 0.510; c/r.t.: cov($k_a$, $k_b$) = 0.586, cov($k_a$, $k_c$) = 1.051, and cov($k_b$, $k_c$) = 0.326. d/10°: cov($k_a$, $k_b$) = -0.525, cov($k_a$, $k_c$) = 0.404, and cov($k_b$, $k_c$) = 0.521; d/r.t.: cov($k_a$, $k_b$) = 0.801, cov($k_a$, $k_c$) = 1.378, and cov($k_b$, $k_c$) = -0.388. e/10°: cov($k_a$, $k_b$) = 0.499, cov($k_a$, $k_c$) = 0.672, and cov($k_b$, $k_c$) = 0.198; e/r.t.: cov($k_a$, $k_b$) = -0.786, cov($k_a$, $k_c$) = -1.029, and cov($k_b$, $k$c) = -0.837.

Table 3. *Kinetic Data for PNA Self-Replication Assays Involving Different Nucleophilic Catalysts*[a]

| Catalyst | $T$ | Model | RMS (%) | $k_a$ (M$^{-1}$ s$^{-1}$) | $k_b$ (M$^{-3/2}$ s$^{-1}$) | $k_c$ (s$^{-1}$) | $\varepsilon$ (M$^{-1/2}$) |
|---|---|---|---|---|---|---|---|
| - | 10° | A | 3.18 | $(4.2\pm0.5)\times10^{-4}$ | - | $(6\pm2)\times10^{-6}$ | - |
|  |  | B | 0.867 | $(1.7\pm0.1)\times10^{-4}$ | $(1.6\pm0.1)\times10^{-2}$ | $(1.0\pm0.8)\times10^{-6}$ | 94 |
|  | r.t. | A | 0.344 | $(9.4\pm0.1)\times10^{-4}$ | - | $(2.18\pm0.04)\times10^{-5}$ | - |
|  |  | B | 0.344 | $(9.4\pm0.5)\times10^{-4}$ | $(0.1\pm4.0)\times10^{-3}$ | $(2.2\pm0.1)\times10^{-5}$ | 0.1 |
| Im | 10° | A | 2.68 | $(4.7\pm0.3)\times10^{-4}$ | - | $(0.2\pm1.3)\times10^{-6}$ | - |
|  |  | B | 1.47 | $(2.5\pm0.2)\times10^{-4}$ | $(9.9\pm0.8)\times10^{-3}$ | $(1.5\pm0.7)\times10^{-6}$ | 40 |
|  | r.t. | A | 1.10 | $(5.5\pm0.2)\times10^{-4}$ | - | $(8\pm1)\times10^{-6}$ | - |
|  |  | B | 1.09 | $(5.2\pm0.2)\times10^{-4}$ | $(3\pm2)\times10^{-3}$ | $(9.0\pm0.8)\times10^{-6}$ | 6 |
| 1-MeIm | 10° | A | 1.22 | $(5.6\pm0.2)\times10^{-4}$ | - | $(5.1\pm0.7)\times10^{-6}$ | - |
|  |  | B | 1.31 | $(5.2\pm0.3)\times10^{-4}$ | $(0.3\pm1.0)\times10^{-3}$ | $(3.7\pm0.9)\times10^{-6}$ | 0.6 |
|  | r.t. | A | 2.13 | $(3.53\pm0.08)\times10^{-3}$ | - | $(0.9\pm4.2)\times10^{-7}$ | - |
|  |  | B | 2.25 | $(1.43\pm0.06)\times10^{-3}$ | $(9.2\pm0.2)\times10^{-2}$ | $(5.3\pm0.4)\times10^{-6}$ | 66 |
| Pyridine | 10° | A | 1.33 | $(5.3\pm0.3)\times10^{-4}$ | - | $(0.04\pm1.15)\times10^{-6}$ | - |
|  |  | B | 0.622 | $(3.7\pm0.1)\times10^{-4}$ | $(6.9\pm0.5)\times10^{-3}$ | $(7\pm5)\times10^{-7}$ | 19 |
|  | r.t. | A | 1.27 | $(4.44\pm0.06)\times10^{-3}$ | - | $(1.4\pm0.4)\times10^{-6}$ | - |
|  |  | B | 1.16 | $(4.32\pm0.05)\times10^{-3}$ | $(6\pm2)\times10^{-3}$ | $(1.9\pm0.4)\times10^{-6}$ | 1.39 |
| HOAt | 10° | A | 7.75 | $(3.7\pm0.2)\times10^{-2}$ | - | $(1.6\pm0.3)\times10^{-5}$ | - |
|  |  | B | 7.75 | $(3.7\pm0.2)\times10^{-2}$ | $(0.1\pm5.4)\times10^{-3}$ | $(1.6\pm0.3)\times10^{-5}$ | 0.003 |
|  | r.t. | A | 4.56 | $(8.7\pm0.3)\times10^{-2}$ | - | $(1.43\pm0.08)\times10^{-4}$ | - |
|  |  | B | 4.58 | $(9\pm1)\times10^{-2}$ | $(5\pm2)\times10^{-2}$ | $(1.4\pm0.2)\times10^{-4}$ | 0.6 |

[a] 5 mM tri-PNAs **A** and **B**, 0.2 M EDC, 0.2 M MOPS pH 7.2, 0.2 M NaCl, 0.1 M catalyst, 6 µl scale

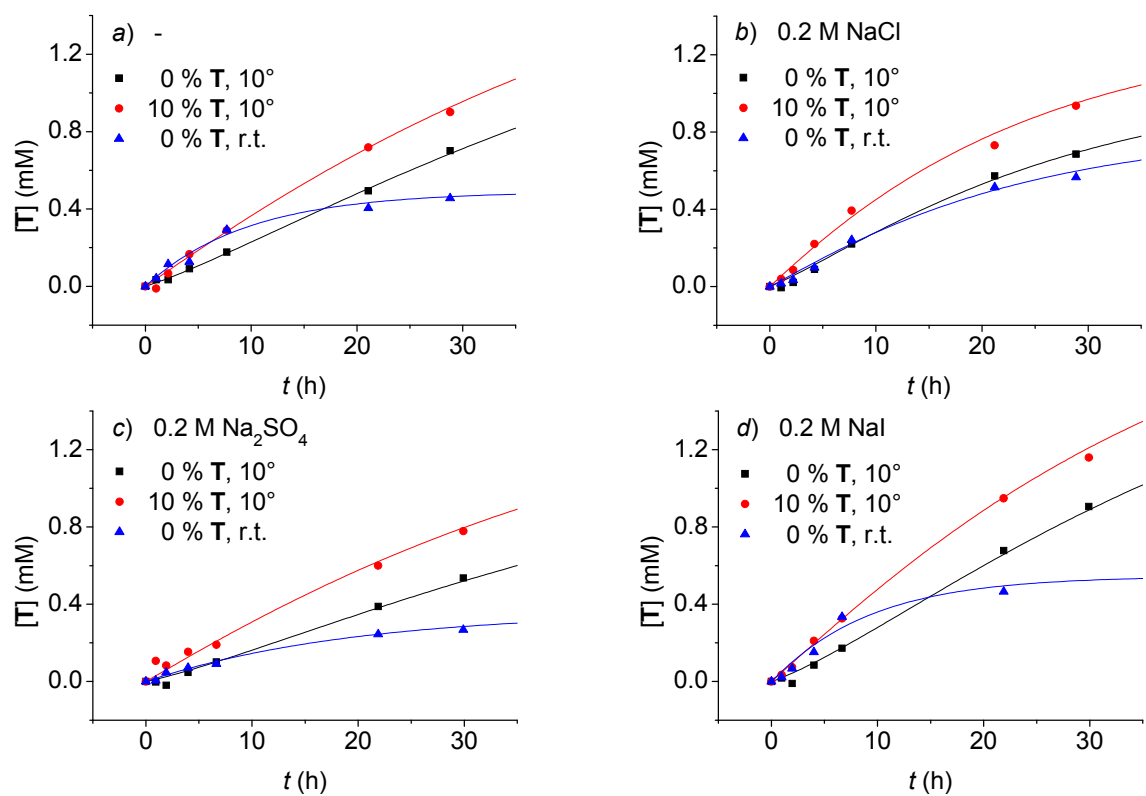

Fig 4. *Kinetics of PNA Self-Replication Assays Involving Different Salts.* Shown are the experimental data points and fittings according to model B. Conditions: 5 mM tri-PNAs **A** and **B**, 0.2 M EDC, 0.2 M MOPS pH 7.6, b: 0.2 M NaCl, c: 0.2 M Na$_2$SO$_4$, d: 0.2 M NaI, 0.1 M imidazole, 6 µl scale. a/10°: cov($k_a$, $k_b$) = -0.436, cov($k_a$, $k_c$) = 0.122, and cov($k_b$, $k_c$) = 0.822; a/r.t.: cov($k_a$, $k_b$) = 0.559, cov($k_a$, $k_c$) = 1.433, and cov($k_b$, $k_c$) = -0.447. b/10°: cov($k_a$, $k_b$) = -0.434, cov($k_a$, $k_c$) = 0.220, and cov($k_b$, $k_c$) = 0.753; b/r.t.: cov($k_a$, $k_b$) = 0.796, cov($k_a$, $k_c$) = 2.132, and cov($k_b$, $k_c$) = -0.767. c/10°: cov($k_a$, $k_b$) = -0.149, cov($k_a$, $k_c$) = 0.469, and cov($k_b$, $k_c$) = 0.772; c/r.t.: cov($k_a$, $k_b$) = 0.610, cov($k_a$, $k_c$) = 1.783, and cov($k_b$, $k_c$) = -0.689. d/10°: cov($k_a$, $k_b$) = -0.369, cov($k_a$, $k_c$) = 0.343, and cov($k_b$, $k_c$) = 0.711; d/r.t.: cov($k_a$, $k_b$) = 0.430, cov($k_a$, $k_c$) = 1.139, and cov($k_b$, $k_c$) = -0.281.

Table 4. *Kinetic Data for PNA Self-Replication Assays Involving Different Salts*[a])

| Salt | $T$ | Model | RMS (%) | $k_a$ (M$^{-1}$ s$^{-1}$) | $k_b$ (M$^{-3/2}$ s$^{-1}$) | $k_c$ (s$^{-1}$) | $\varepsilon$ (M$^{-1/2}$) |
|---|---|---|---|---|---|---|---|
| - | 10° | A | 2.65 | (3.7±0.4)×10$^{-4}$ | - | (0.06±2.02)×10$^{-6}$ | - |
|  |  | B | 0.674 | (1.73±0.08)×10$^{-4}$ | (1.11±0.07)×10$^{-2}$ | (2.9±0.6)×10$^{-6}$ | 64.2 |
|  | r.t. | A | 0.928 | (5.7±0.3)×10$^{-4}$ | - | (2.5±0.2)×10$^{-5}$ | - |
|  |  | B | 0.927 | (5.6±0.3)×10$^{-4}$ | (1±2)×10$^{-3}$ | (2.5±0.2)×10$^{-5}$ | 2 |
| NaCl | 10° | A | 3.02 | (4.6±0.5)×10$^{-4}$ | - | (3±2)×10$^{-6}$ | - |
|  |  | B | 0.949 | (2.2±0.1)×10$^{-4}$ | (1.60±0.09)×10$^{-2}$ | (8.8±0.8)×10$^{-6}$ | 73 |
|  | r.t. | A | 1.12 | (3.7±0.3)×10$^{-4}$ | - | (6±2)×10$^{-6}$ | - |
|  |  | B | 0.973 | (2.9±0.2)×10$^{-4}$ | (1.0±0.2)×10$^{-2}$ | (1.1±0.1)×10$^{-5}$ | 34 |
| Na$_2$SO$_4$ | 10° | A | 3.03 | (2.8±0.5)×10$^{-4}$ | - | (0.03±2.96)×10$^{-6}$ | - |
|  |  | B | 1.10 | (1.1±0.1)×10$^{-4}$ | (1.11±0.07)×10$^{-2}$ | (4±1)×10$^{-6}$ | 100 |
|  | r.t. | A | 0.409 | (2.1±0.1)×10$^{-4}$ | - | (1.3±0.1)×10$^{-5}$ | - |
|  |  | B | 0.407 | (2.0±0.1)×10$^{-4}$ | (1±1)×10$^{-3}$ | (1.3±0.1)×10$^{-5}$ | 5 |
| NaI | 10° | A | 3.54 | (5.0±0.6)×10$^{-4}$ | - | (0.01±2.10)×10$^{-6}$ | - |
|  |  | B | 0.781 | (1.75±0.09)×10$^{-4}$ | (1.66±0.05)×10$^{-2}$ | (3.5±0.4)×10$^{-6}$ | 94.9 |
|  | r.t. | A | 1.55 | (6.6±0.6)×10$^{-4}$ | - | (2.5±0.4)×10$^{-5}$ | - |
|  |  | B | 1.51 | (6.0±0.6)×10$^{-4}$ | (7±3)×10$^{-3}$ | (2.7±0.4)×10$^{-5}$ | 12 |

[a]) 5 mM tri-PNAs **A** and **B**, 0.2 M EDC, 0.2 M MOPS pH 7.6, 0.2 M salt, 0.1 M imidazole, 6 μl scale

*Figure 5* and *Table 5* show that the yield of the ligation product is limited by the hydrolysis of the condensing agent: Increasing the EDC concentration from 0.2 to 0.4 M doubles the yield of hexa-PNA **T**, while not biasing the autocatalytic excess factor $\varepsilon$. Further, rate and yield of the ligation increase with decreasing pH, as can be expected for an EDC mediated peptide coupling [59, 60] (*Fig. 6*). In terms of autocatalytic efficiency, pH 6.6 was the best of the three pH values tested ($\varepsilon$ = 238 M$^{-1/2}$, *Table 6*). Subsequently, the combination of pH 6.6 and 0.4 M EDC allowed to establish reaction conditions optimized for fast replication and high conversion (*Fig. 7, Table 7*).

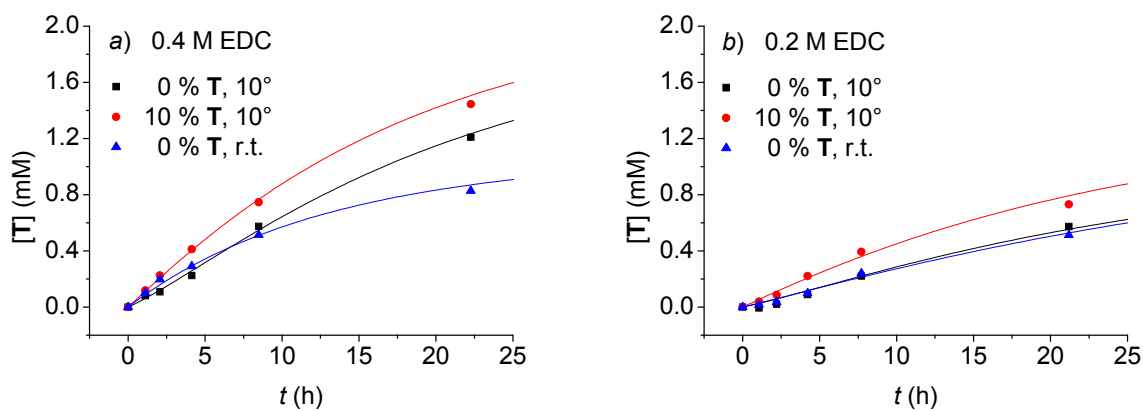

Fig. 5. *Effect of EDC Concentration on the Kinetics of the PNA Self-Replication System.* Shown are the experimental data points and fittings according to model B. Conditions: 5 mM tri-PNAs **A** and **B**, a: 0.4 M EDC, b: 0.2 M EDC, 0.2 M MOPS pH 7.6, 0.2 M NaCl, 6 μl scale. a/10°: cov($k_a$, $k_b$) = -0.753, cov($k_a$, $k_c$) = -0.130, and cov($k_b$, $k_c$) = 0.711; a/r.t.: cov($k_a$, $k_b$) = -0.320, cov($k_a$, $k_c$) = 0.774, and cov($k_b$, $k_c$) = 0.216. b/10°: cov($k_a$, $k_b$) = -0.363, cov($k_a$, $k_c$) = 0.512, and cov($k_b$, $k_c$) = 0.533; b/r.t.: cov($k_a$, $k_b$) = 0.360, cov($k_a$, $k_c$) = 1.283, and cov($k_b$, $k_c$) = -0.500.

Table 5. *Kinetic Data for PNA Self-Replication Assays Involving Different EDC Concentrations*[a])

| [EDC] (M) | $T$ | Model | RMS (%) | $k_a$ (M$^{-1}$ s$^{-1}$) | $k_b$ (M$^{-3/2}$ s$^{-1}$) | $k_c$ (s$^{-1}$) | $\varepsilon$ (M$^{-1/2}$) |
|---|---|---|---|---|---|---|---|
| 0.4 | 10° | A | 3.18 | (9.9±0.7)×10$^{-4}$ | - | (1±2)×10$^{-6}$ | - |
|  |  | B | 0.674 | (4.6±0.2)×10$^{-4}$ | (3.0±0.1)×10$^{-2}$ | (8.0±0.4)×10$^{-6}$ | 65 |
|  | r.t. | A | 0.585 | (1.02±0.02)×10$^{-3}$ | - | (1.76±0.09)×10$^{-5}$ | - |
|  |  | B | 0.784 | (9.4±0.3)×10$^{-4}$ | (0.07±9.14)×10$^{-4}$ | (1.5±0.1)×10$^{-5}$ | 0.007 |
| 0.2 | 10° | A | 2.53 | (4.5±0.5)×10$^{-4}$ | - | (2±3)×10$^{-6}$ | - |
|  |  | B | 1.02 | (2.3±0.2)×10$^{-4}$ | (1.59±0.08)×10$^{-2}$ | (9±1)×10$^{-6}$ | 69 |
|  | r.t. | A | 0.884 | (3.3±0.2)×10$^{-4}$ | - | (1±1)×10$^{-6}$ | - |
|  |  | B | 0.811 | (2.7±0.2)×10$^{-4}$ | (8±1)×10$^{-3}$ | (6±1)×10$^{-6}$ | 27 |

[a]) 5 mM tri-PNAs **A** and **B**, 0.2 M MOPS pH 7.6, 0.2 M NaCl, 0.1 M imidazole, 6 μl scale

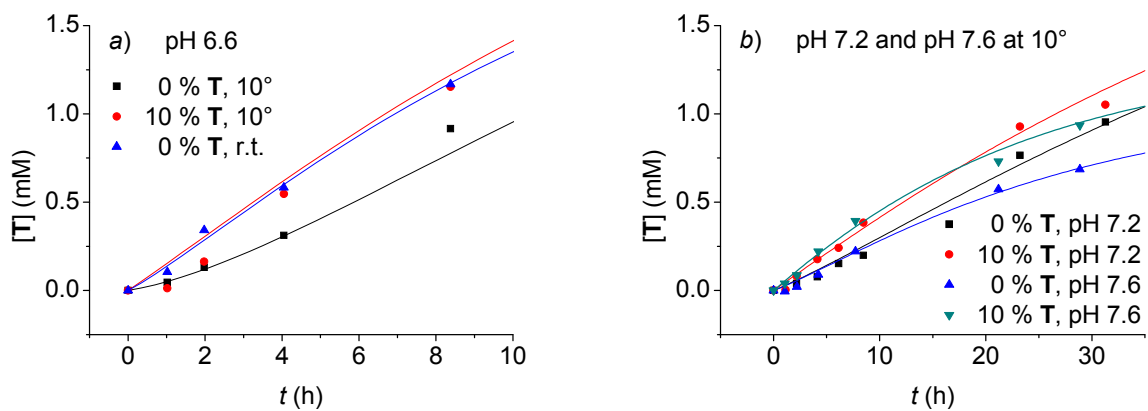

Fig. 6. *Effect of Different pH Values on the Kinetics of the PNA Self-Replication System.* Shown are the experimental data points and fittings according to model B. Conditions: 5 mM tri-PNAs **A** and **B**, 0.2 M EDC, 0.2 M MOPS, 0.2 M NaCl, 0.1 M imidazole, 6 μl scale. a/10°: cov($k_a$, $k_b$) = 0.033, cov($k_a$, $k_c$) = 0.907, and cov($k_b$, $k_c$) = -0.322; a/r.t.: cov($k_a$, $k_b$) = 0.360, cov($k_a$, $k_c$) = 0.151, and cov($k_b$, $k_c$) = -0.345. b/pH 7.2: cov($k_a$, $k_b$) = -0.492, cov($k_a$, $k_c$) = 0.355, and cov($k_b$, $k_c$) = 0.596; b/pH 7.6: cov($k_a$, $k_b$) = 0.220, cov($k_a$, $k_c$) = -0.434, and cov($k_b$, $k_c$) = 0.753.

Table 6. *Kinetic Data for the PNA Self-Replicating System under Different pH Conditions*[a])

| pH | $T$ | Model | RMS (%) | $k_a$ (M$^{-1}$ s$^{-1}$) | $k_b$ (M$^{-3/2}$ s$^{-1}$) | $k_c$ (s$^{-1}$) | $\varepsilon$ (M$^{-1/2}$) |
|---|---|---|---|---|---|---|---|
| 6.6 | 10° | A | - | - | - | - | - |
|  |  | B | 3.33 | (2.6±0.8)×10$^{-4}$ | (6.2±0.1)×10$^{-2}$ | (5±3)×10$^{-6}$ | 238 |
|  | r.t. | A | 1.83 | (2.0±0.1)×10$^{-3}$ | - | (0.05±4.76)×10$^{-6}$ | - |
|  |  | B | 1.37 | (1.15±0.08)×10$^{-3}$ | (5.3±0.3)×10$^{-2}$ | (9±3)×10$^{-6}$ | 35 |
| 7.2 | 10° | A | 2.68 | (4.7±0.3)×10$^{-4}$ | - | (0.2±1.3)×10$^{-6}$ | - |
|  |  | B | 1.47 | (2.5±0.2)×10$^{-4}$ | (9.9±0.8)×10$^{-3}$ | (1.5±0.7)×10$^{-6}$ | 40 |
|  | r.t. | A | 1.10 | (5.5±0.2)×10$^{-4}$ | - | (8±1)×10$^{-6}$ | - |
|  |  | B | 1.09 | (5.2±0.2)×10$^{-4}$ | (3±2)×10$^{-3}$ | (9.6±0.8)×10$^{-6}$ | 6 |
| 7.6 | 10° | A | 3.02 | (4.6±0.5)×10$^{-4}$ | - | (3±2)×10$^{-6}$ | - |
|  |  | B | 0.949 | (2.2±0.1)×10$^{-4}$ | (1.60±0.09)×10$^{-2}$ | (8.8±0.8)×10$^{-6}$ | 73 |
|  | r.t. | A | 1.12 | (3.7±0.3)×10$^{-4}$ | - | (6±2)×10$^{-6}$ | - |
|  |  | B | 0.973 | (2.9±0.2)×10$^{-4}$ | (1.0±0.2)×10$^{-2}$ | (1.1±0.1)×10$^{-5}$ | 34 |

a) 5 mM tri-PNAs **A** and **B**, 0.2 M EDC, 0.2 M MOPS, 0.2 M NaCl, 0.1 M imidazole, 6 μl scale

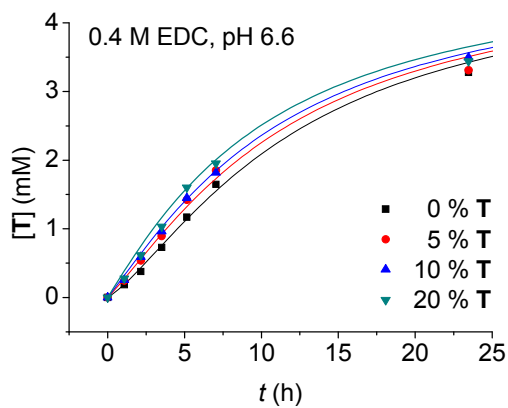

Fig. 7. *Kinetics of the PNA Self-Replicating System under Conditions Optimized for Fast Replication and High Conversion.* Shown are the experimental data points and fittings according to model B. Conditions: 5 mM tri-PNAs **A** and **B**, 0.4 M EDC, 0.2 M MOPS pH 6.6, 0.1 M imidazole, 10°, 6 µl scale. cov($k_a$, $k_b$) = -0.621, cov($k_a$, $k_c$) = 0.317, and cov($k_b$, $k_c$) = 0.242.

Table 7. *Kinetic Data for the PNA Self-Replicating System under Conditions Optimized for Fast Replication and High Conversion*[a])

| Model | RMS (%) | $k_a$ (M$^{-1}$ s$^{-1}$) | $k_b$ (M$^{-3/2}$ s$^{-1}$) | $k_c$ (s$^{-1}$) | $\varepsilon$ (M$^{-1/2}$) |
|---|---|---|---|---|---|
| A | 7.43 | (4.3±0.2)×10$^{-3}$ | - | (0.2±1.2)×10$^{-6}$ | - |
| B | 2.55 | (1.21±0.05)×10$^{-3}$ | (9.0±0.1)×10$^{-2}$ | (0.02±2.76)×10$^{-7}$ | 75 |

[a]) 5 mM tri-PNAs **A** and **B**, 0.4 M EDC, 0.2 M MOPS pH 6.6, 0.1 M imidazole, 10°, 6 µl

Increasing the imidazole concentration from 0.1 to 0.2 M gave no significant change in the autocatalytic excess factor *ε*, while slightly reducing the rate of the reaction (*Fig. 8, Table 8*). The latter can be attributed to the influence of additional basic imidazole on the pH value.

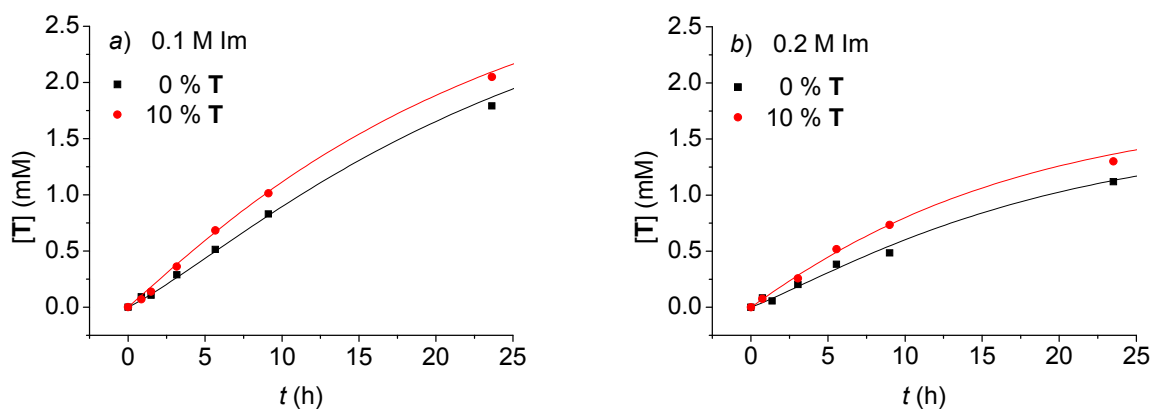

Fig 8. *Kinetics of PNA Self-Replication Assays Involving Different Imidazole Concentrations.* Shown are the experimental data points and fittings according to model B. Conditions: 5 mM tri-PNAs **A** and **B**, 0.2 M EDC, 0.2 M MOPS pH 7.2, 0.2 M NaI, a: 0.1 M imidazole, b: 0.2 M imidazole, 10°, 6 µl scale. a: cov($k_a$, $k_b$) = -0.903, cov($k_a$, $k_c$) = -0.556, and cov($k_b$, $k_c$) = 0.834. b: cov($k_a$, $k_b$) = -0.885, cov($k_a$, $k_c$) = -0.504, and cov($k_b$, $k_c$) = 0.828.

Table 8. *Kinetic Data for the PNA Self-Replicating System under Different Imidazole Concentrations*[a])

| [Im] (mM) | Model | RMS (%) | $k_a$ (M$^{-1}$ s$^{-1}$) | $k_b$ (M$^{-3/2}$ s$^{-1}$) | $k_c$ (s$^{-1}$) | $\varepsilon$ (M$^{-1/2}$) |
|---|---|---|---|---|---|---|
| 0.1 | A | 3.56 | (1.48±0.07)×10$^{-3}$ | - | (0.3±1.2)×10$^{-6}$ | - |
|  | B | 0.920 | (6.6±0.3)×10$^{-4}$ | (3.1±0.2)×10$^{-2}$ | (2.2±0.4)×10$^{-6}$ | 47 |
| 0.2 | A | 2.97 | (9.6±0.6)×10$^{-4}$ | - | (4±2)×10$^{-6}$ | - |
|  | B | 1.16 | (5.0±0.4)×10$^{-4}$ | (2.6±0.3)×10$^{-2}$ | (1.0±0.1)×10$^{-5}$ | 52 |

[a]) 5 mM tri-PNAs **A** and **B**, 0.2 M EDC, 0.2 M MOPS pH 7.2, 0.2 M NaI, 10°, 6 µl scale

In the following, we explored the self-replicating system in the eutectic phase system as it is known to favor aggregates and organized structures which in turn are a prerequisite for replication [61-66]. In addition, hydrolysis is repressed under these conditions due to a lower water activity. *Figure 9* indicates that added template has only a small effect on the rate and yield of the ligation, a finding which is supported by the kinetic data when compared to the analogous experiment at 10° (*Table 9*). This can be explained by pronounced product inhibition that might overwhelm an assumed positive effect on the termolecular complex at temperatures far below the melting point of the product duplex.

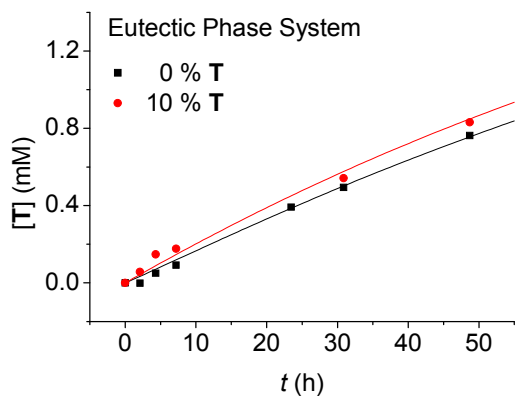

Fig. 9. *Kinetics of PNA Self-Replication in the Eutectic Phase System.* Shown are the experimental data points and fittings according to model B. Conditions: 5 mM tri-PNAs **A** and **B**, 0.2 M EDC, 0.2 M MOPS pH 7.6, 0.2 M NaCl, 0.1 M imidazole, -19°, 12 µl scale. cov($k_a$, $k_b$) = -0.398, cov($k_a$, $k_c$) = 0.382, and cov($k_b$, $k_c$) = 0.669.

Table 9. *Kinetic Data for PNA Self-Replication in the Eutectic Phase at -19° and in the Liquid Phase at 10°*[a])

| $T$ | Model | RMS (%) | $k_a$ (M$^{-1}$ s$^{-1}$) | $k_b$ (M$^{-3/2}$ s$^{-1}$) | $k_c$ (s$^{-1}$) | $\varepsilon$ (M$^{-1/2}$) |
|---|---|---|---|---|---|---|
| -19° | A | 1.46 | (2.1±0.1)×10$^{-4}$ | - | (0.5±7.1)×10$^{-7}$ | - |
|  | B | 1.02 | (1.70±0.09)×10$^{-4}$ | (2.7±0.5)×10$^{-3}$ | (9±6)×10$^{-7}$ | 16 |
| 10° | A | 2.53 | (4.5±0.5)×10$^{-4}$ | - | (2±3)×10$^{-6}$ | - |
|  | B | 1.02 | (2.3±0.2)×10$^{-4}$ | (1.59±0.08)×10$^{-2}$ | (9±1)×10$^{-6}$ | 69 |

[a]) 5 mM tri-PNAs **A** and **B**, 0.2 M EDC, 0.2 M MOPS pH 7.6, 0.2 M NaCl, 0.1 M imidazole

Finally, the influence of uncharged co-solvents on the self-replicating system was studied because they were considered to have an impact on the structures and stabilities of the complexes involved [67-70] (*Fig. 10, Table 10*). In fact, both, the addition of PEG3350 and PEG400 improved the template effect: While 20 % PEG3350 increased the autocatalytic excess factor $\varepsilon$ from 69 to 126 $M^{-1/2}$, 20 % PEG400 lead to the attainment of the overall maximum value of $\varepsilon$ for the experiments presented ($\varepsilon = 324$ $M^{-1/2}$). Surprisingly, the ligation yield was amplified in the presence of PEG3350 but reduced by a factor of two to three in the presence of PEG400 when compared to the corresponding experiment without added co-solvent (*Fig. 4b, Table 10*). Additionally, the amount of side-products increased in the presence of both PEGs (data not shown).

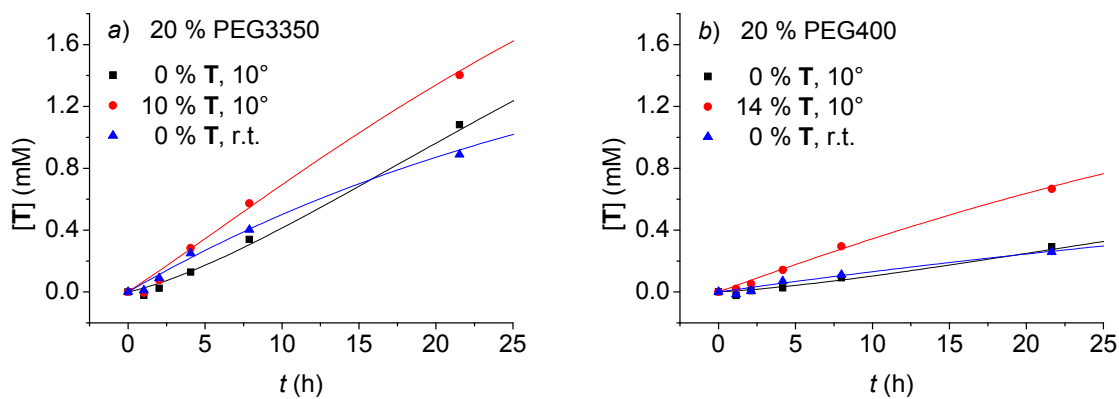

Fig. 10. *Effect of PEG on the Kinetics of the PNA Self-Replicating System.* Shown are the experimental data points and fittings according to model B. Conditions: 5 mM tri-PNAs **A** and **B**, 0.2 M EDC, 0.2 M MOPS pH 7.6, 0.2 M NaCl, 0.1 M imidazole, a: 20 % PEG3350, b: 20 % PEG400, 6 μl scale. a/10°: cov($k_a$, $k_b$) = -0.870, cov($k_a$, $k_c$) = -0.666, and cov($k_b$, $k_c$) = 0.938; a/r.t.: cov($k_a$, $k_b$) = 0.225, cov($k_a$, $k_c$) = 1.071, and cov($k_b$, $k_c$) = -0.328. b/10°: cov($k_a$, $k_b$) = -0.167, cov($k_a$, $k_c$) = 0.224, and cov($k_b$, $k_c$) = 0.889; b/r.t.: cov($k_a$, $k_b$) = 0.304, cov($k_a$, $k_c$) = 1.219, and cov($k_b$, $k$c) = -0.444.

Table 10. *Kinetic Data for the PNA Self-Replicating System in the Absence and Presence of PEG*[a])

| PEG | $T$ | Fit | RMS (%) | $k_a$ (M$^{-1}$ s$^{-1}$) | $k_b$ (M$^{-3/2}$ s$^{-1}$) | $k_c$ (s$^{-1}$) | $\varepsilon$ (M$^{-1/2}$) |
|---|---|---|---|---|---|---|---|
| - | 10° | A | 2.53 | (4.5±0.5)×10$^{-4}$ | - | (2±3)×10$^{-6}$ | - |
|  |  | B | 1.02 | (2.3±0.2)×10$^{-4}$ | (1.59±0.08)×10$^{-2}$ | (9±1)×10$^{-6}$ | 69 |
|  | r.t. | A | 0.884 | (3.3±0.2)×10$^{-4}$ | - | (1±1)×10$^{-6}$ | - |
|  |  | B | 0.811 | (2.7±0.2)×10$^{-4}$ | (8±1)×10$^{-3}$ | (6±1)×10$^{-6}$ | 27 |
| 400 | 10° | A | 3.79 | (2.7±0.7)×10$^{-4}$ | - | (0.09±6.08)×10$^{-6}$ | - |
|  |  | B | 0.604 | (4.2±0.5)×10$^{-5}$ | (1.36±0.06)×10$^{-2}$ | (4±1)×10$^{-6}$ | 324 |
|  | r.t. | A | 0.757 | (1.5±0.2)×10$^{-4}$ | - | (2±3)×10$^{-6}$ | - |
|  |  | B | 0.754 | (1.5±0.2)×10$^{-4}$ | (0.7±1.3)×10$^{-3}$ | (3±3)×10$^{-6}$ | 5 |
| 3350 | 10° | A | 5.02 | (8±1)×10$^{-4}$ | - | (1±3)×10$^{-6}$ | - |
|  |  | B | 1.45 | (1.9±0.4)×10$^{-4}$ | (2.4±0.3)×10$^{-2}$ | (0.02±1.56)×10$^{-6}$ | 126 |
|  | r.t. | A | 1.10 | (6.5±0.3)×10$^{-4}$ | - | (3±1)×10$^{-6}$ | - |
|  |  | B | 1.08 | (6.2±0.3)×10$^{-4}$ | (3.1±0.9)×10$^{-3}$ | (4±2)×10$^{-6}$ | 5 |

[a]) 5 mM tri-PNAs **A** and **B**, 0.2 M EDC, 0.2 M MOPS pH 7.6, 0.2 M NaCl, 0.1 M imidazole, 20 % PEG, 6 μl scale

**4. Conclusions and Outlook.** – In conclusion, we presented the first self-replicating system based on peptide nucleic acids in which a hexa-PNA catalyzes its own formation from two trimeric building blocks. The kinetics of the reaction were followed by RP-HPLC and analyzed by our *SimFit* program, thereby revealing parabolic growth according to the square-root law. The study concentrated on optimizing the autocatalytic efficiency of the system by systematically varying the reaction conditions for the EDC mediated condensation. Thus, we were able to increase the autocatalytic excess factor $\varepsilon$ by two orders of magnitude up to 324 M$^{-1/2}$. Moreover, we found conditions for efficient and template free ligation of PNAs in aqueous solutions. Former methods used native chemical ligation or the derived native chemical *i*Cys-ligation and necessitate the preparation of particular building blocks [35, 36, 38, 71]. We are currently continuing the screening of nucleophilic catalysts and uncharged co-solvents to enhance the autocatalytic efficiency of the system. In addition, HPLC monitoring will be improved by synthesizing a chemically marked template in order to avoid errors during the calculations of differences between HPLC integrals. In the long term, the analysis of the system will be expanded to kinetic $^{19}$F-NMR titration.

# Experimental Part

*General*. The synthesis of F-PNA **A** and PNA **B** was reported elsewhere [41]. Hexa-PNA **T** was isolated from preliminary experiments without initially added template using semi-prep. RP-HPLC. MALDI-MS: 1903.197 (95, $[M + 3\,H]^+$). HR-ESI-MS: 1904.8704 (5, $[M + 5\,H]^+$, $C_{79}H_{112}F_2N_{37}O_{18}^+$, calc.: 1904.8954); 1903.8673 (15, $[M + 4\,H]^+$, $C_{79}H_{111}F_2N_{37}O_{18}^+$, calc.: 1903.8875); 1902.8653 (41, $[M + 3\,H]^+$, $C_{79}H_{110}F_2N_{37}O_{18}^+$, calc.: 1902.8797); 1901.8629 (95, $[M + 2\,H]^+$, $C_{79}H_{109}F_2N_{37}O_{18}^+$, calc.: 1901.8719); 1900.8602 (100, $[M + H]^+$, $C_{79}H_{108}F_2N_{37}O_{18}^+$, calc.: 1900.8641). All other chemicals were used as delivered, usually in p.a. quality. Temperatures are given in degrees Celsius. Freeze drying: *Christ Alpha 1-2*. HPLC: *Beckman Gold* (126 Solvent Module, 168 Detector) with *Jetstream 2 Plus Column-Thermostat*; *A*: 0.1% TFA in deionized and dist. H$_2$O, B: 0.1% TFA in MeCN; $T_{column}$, 55°. Anal. HPLC: 250×4.6 mm *Supelco Ascentis RP-Amide* 5 μm; flow, 1 ml/min. Semi-perp. HPLC: 250 × 10 mm *Supelco Discovery BIO Wide Pore C18* 5 μm; flow, 3 ml/min. MS: *Thermo Scientific LTQ-Orbitrap XL* (HR-ESI), *Bruker daltonics autoflex* (MALDI-TOF; matrix: 2',4',6'-trihydroxyacetophe-none). UV: *Varian Cary 1E*.

*Abbreviations*. EDC: 1-ethyl-3-[3-(dimethylamino)propyl]carbodiimide, EDU: 1-ethyl-3-[3-(dimethyl-amino)propyl]urea, ESI: electrospray ionization, HOAt: 1-hydroxy-7-azabenzotriazole, Im: imidazole, MALDI-TOF: matrix-assisted laser-desorption ionization time of flight, 1-MeIm: 1-methylimidazole, MOPS: 3-(*N*-morpholino)propanesulfonic acid, PEG: polyethylene glycol, RMS: root mean square, TFA: trifluoroacetic acid.

*Ligation/Self-Replication-Assays*. Reaction solutions were prepared in 500 μl polypropylene sample tubes (*Sarstedt*) from a 10 mM stock solution of tri-PNAs **A** and **B**. To start the reaction, the same volume of a solution of EDC, nucleophilic catalyst, and salt in the respective buffer was added. The mixtures were immediately vortexed, centrifuged, and split into a series of sample tubes which, where required, contained lyophilized hexa-PNA **T**. The resulting solutions were cooled (*MJ Research MiniCycler PTC-150*) or kept at r.t. At appropriate times, 0.5 μl reaction solution was added to 500 μl of an aqueous solution containing 5 % MeCN and 0.1 % TFA. After mixing and vortexing, the resulting solution was immediately analyzed by RP-HPLC or kept at r.t. Storage at r.t. for several days had no effect on sample composition.

# Supplement

## 1 Analytical Data for Hexa-PNA T

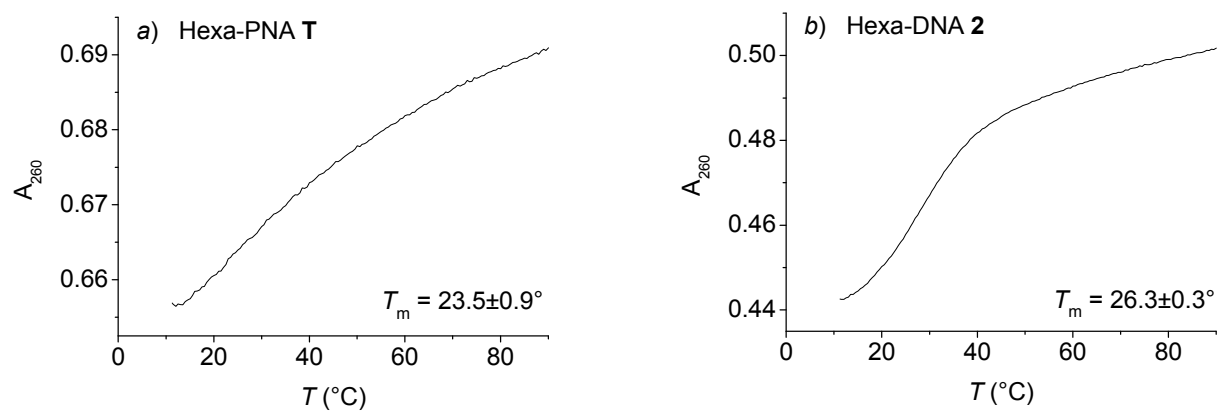

**S1:** *UV Melting Profiles at 260 nm.* a) Hexa-PNA **T** ($^N$cfgcag$^C$). b) Comparable hexa-DNA **2** (5'-CTGCAG-3'). Conditions: 0.25 M sodium phosphate pH 7, 0.1 M NaCl, 1°/min.

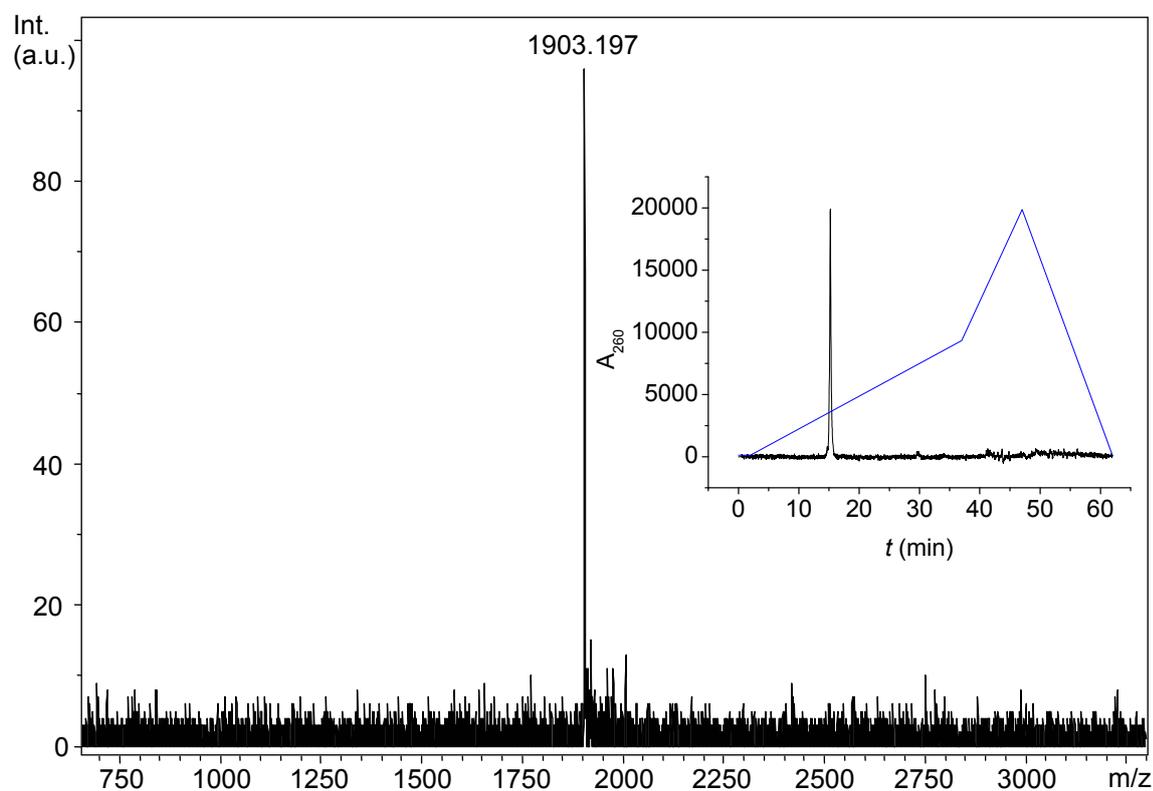

**S2.** *MALDI-TOF Mass Spectra and RP-HPLC Plot of Hexa-PNA* **T** (therein). Conditions: LP, THAP (MS); 55°, 1 ml/min (HPLC). Gradient: 2 min 5 %B, 5 → 40 %B in 35 min, 40 → 80 %B in 10 min, 80 %B → 5 %B in 15 min.

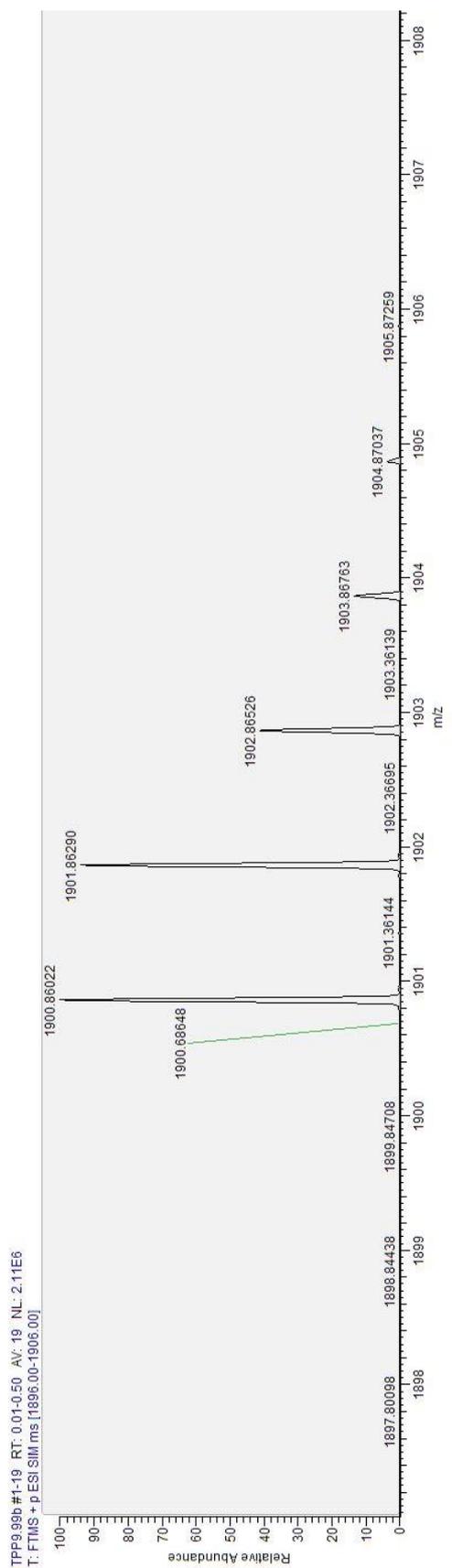

**S3.** *HR-ESI-MS Spectra of Hexa-PNA* **T**.

## 2 HPLC Calibration and Analysis

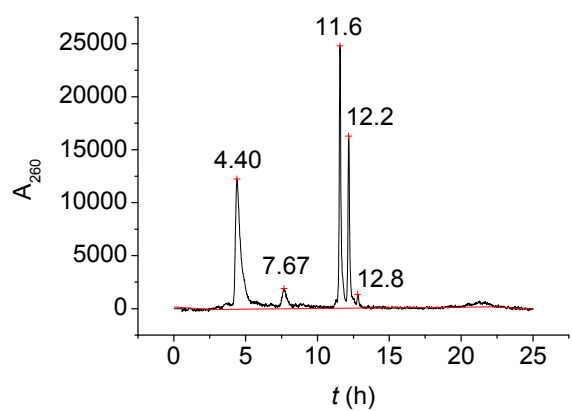

**S4.** *Representative RP-HPLC Plot.* The peaks were isolated and analyzed by MALDI-TOF-MS. $t_r$ = 4.40 min: **B**; $t_r$ = 7.67 min **B**+EDC; $t_r$ = 11.6 min: **T**; $t_r$ = 12.2 min: **A**; 12.8 min **A**+EDC.

Concentrations of PNA **A** and **T** were calculated from the respective HPLC areas by means of an internal calibration because both compounds gave sharp signals with comparable retention times (*S4.*). To this end, a calibration curve was recorded that correlated the area ratio with the concentration ratio of both compounds (*S5.*). Thus, the results are independent of changes in sample concentration or volume errors during sampling, dilution, and injection. Empirically, we found a linear relationship (*eq. i*).

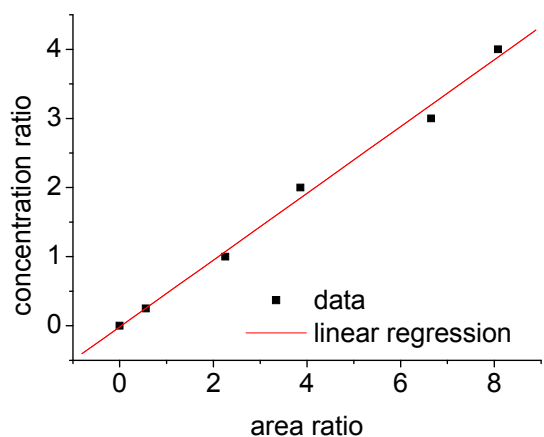

**S5.** *Calibration Curve for Converting the Area Ratio into the Corresponding Concentration Ratio.* Fit parameter: $R^2$ = 0.9936, SD = 0.1417.

f(*x*) = (-0.02±0.09) + (0.48±0.02) × *x*. (i)

Assuming that tri-PNA **A** reacts solely to hexa-PNA **T** gives:

$$[\mathbf{T}] + [\mathbf{A}] = [\mathbf{T}]_0 + [\mathbf{A}]_0 \tag{ii}$$

[**T**]   =   concentration of **T** at time t

[**A**]   =   concentration of **A** at time t

[**T**]$_0$   =   initial concentration of **T**

[**A**]$_0$   =   initial concentration of **A**

Thus, the concentration ratio κ can be expressed by:

$$\kappa = \frac{[\mathbf{T}]}{[\mathbf{A}]} = \frac{[\mathbf{T}]}{[\mathbf{A}]_0 + [\mathbf{T}]_0 - [\mathbf{T}]} \tag{iii}$$

Finally, we may rewrite *eq. (iii)* to calculate the concentration of **T** at time t:

$$[\mathbf{T}] = \frac{\kappa([\mathbf{T}]_0 + [\mathbf{A}]_0)}{\kappa + 1} \tag{iv}$$

# 3 SimFit Files

## 3.1 SimFit Command Files

### *3.1.1 A-Model*

```
* 0.2 M MOPS pH 7.6, 10°, 0.1 M imidazole, 0.2 M NaI, Table 4

define (1, t, p, 12)  scale (3, 1.00000E+0)
define (2, a, e, 11)  scale (3, 1.00000E+0)
select (t, a)
read (NaI.txt)
choose (exp1, exp2)

reaction (A + B --> T) constant (1, 1E-4,1, 1.00000E+0, 1.00000E-10, 1.00000E+10)
reaction (A --> X) constant (2, 1E-6,2, 1.00000E+0, 1.00000E-10, 1.00000E+10)

constant (show)
reaction (compile)
reaction (show)
int (show)

assign (obs,  a = A + X)
assign (obs,  t = T)
assign (spec, A = a)
assign (spec, B = # 5e-3)
assign (spec, T = t)
assign (spec, X = # 0)

assign (show)

time (h)
conc (mM)
win (0, 35, 5, 0.01, 0, 6, 1, 0.01)
dim (2)
integ (stiff, .0000001, 4, .050000000745058, 100, 10)
simplex (plot)
opar (1e+25)
newton (plot)
plot (spec, file)
make
```

*3.1.2 B-Model*

```
* 0.2 M MOPS pH 7.6, 10°, 0.1 M imidazole, 0.2 M NaI, Fig. 4d , Table 4

define (1, t, p, 12)  scale (3, 1.00000E+0)
define (2, a, e, 11)  scale (3, 1.00000E+0)
select (t, a)
read (NaI.txt)
choose (exp1, exp2)

reaction (         A + B --> T)  constant (1, 1E-4,1, 1.00000E+0, 1.00000E-10, 1.00000E+10)
reaction (A + B + 0.5 T --> 1.5 T)  constant (2, 1E-2,2, 1.00000E+0, 1.00000E-10, 1.00000E+10)
reaction (           A --> X)  constant (3, 1E-6,3, 1.00000E+0, 1.00000E-10, 1.00000E+10)

constant (show)
reaction (compile)
reaction (show)
int (show)

assign (obs,  a = A + X)
assign (obs,  t = T)
assign (spec, A = a)
assign (spec, B = # 5e-3)
assign (spec, T = t)
assign (spec, X = # 0)

assign (show)

time (h)
conc (mM)
win (0, 35, 5, 0.01, 0, 6, 1, 0.01)
dim (3)
integ (stiff, .0000001, 4, .05, 100, 10)
simplex (plot)
opar (1e+25)
newton (plot)
plot (spec, file)
make
```

## 3.2 Concentration-Time Data

### 3.2.1 Replication Assays Using 1-Methylimidazole Buffer

Conditions: 5 mM tri-PNAs **A** and **B**, 0.2 M EDC, 0.4 M 1-methylimidazole pH 7.5, 25 µl scale (see *Fig. 1* and *Table 1*).

$T = 0°$

| numb. | exp. | time[h] | t-t0     | a        | t         |
|-------|------|---------|----------|----------|-----------|
| 1     | 1    | 0.00    | 0.000000 | 0.005000 | 0.0000001 |
| 2     | 1    | 4.25    | 0.000065 | 0.004935 | 0.000065  |
| 3     | 1    | 8.01    | 0.000166 | 0.004834 | 0.000166  |
| 4     | 1    | 22.72   | 0.000503 | 0.004497 | 0.000503  |
| 5     | 1    | 30.69   | 0.000537 | 0.004463 | 0.000537  |
| 6     | 1    | 47.77   | 0.000719 | 0.004281 | 0.000719  |
| 7     | 1    | 54.93   | 0.000786 | 0.004214 | 0.000786  |
| 8     | 1    | 71.18   | 0.000937 | 0.004063 | 0.000937  |
| 9     | 1    | 94.73   | 0.001168 | 0.003832 | 0.001168  |
| 1     | 2    | 0.00    | 0.000000 | 0.005000 | 0.001000  |
| 2     | 2    | 4.25    | 0.000085 | 0.004915 | 0.001085  |
| 3     | 2    | 8.01    | 0.000182 | 0.004818 | 0.001182  |
| 4     | 2    | 22.72   | 0.000447 | 0.004553 | 0.001447  |
| 5     | 2    | 30.69   | 0.000550 | 0.004450 | 0.001550  |
| 6     | 2    | 47.77   | 0.000771 | 0.004229 | 0.001771  |
| 7     | 2    | 54.93   | 0.000797 | 0.004203 | 0.001797  |
| 8     | 2    | 71.18   | 0.000964 | 0.004036 | 0.001964  |
| 9     | 2    | 94.73   | 0.001182 | 0.003818 | 0.002182  |
| 1     | 3    | 0.00    | 0.000000 | 0.005000 | 0.002000  |
| 2     | 3    | 4.25    | 0.000121 | 0.004879 | 0.002121  |
| 3     | 3    | 8.01    | 0.000189 | 0.004811 | 0.002189  |
| 4     | 3    | 22.72   | 0.000470 | 0.004530 | 0.002470  |
| 5     | 3    | 30.69   | 0.000515 | 0.004485 | 0.002515  |
| 6     | 3    | 47.77   | 0.000741 | 0.004259 | 0.002741  |
| 7     | 3    | 54.93   | 0.000823 | 0.004177 | 0.002823  |
| 8     | 3    | 71.18   | 0.000955 | 0.004045 | 0.002955  |
| 9     | 3    | 94.73   | 0.001029 | 0.003971 | 0.003029  |
| end   |      |         |          |          |           |

$T = 10°$

| numb. | exp. | time[h] | t-t0 | a | t |
|---|---|---|---|---|---|
| 1 | 1 | 0.00 | 0.000000 | 0.005000 | 0.0000001 |
| 2 | 1 | 4.15 | 0.000201 | 0.004799 | 0.000201 |
| 3 | 1 | 8.06 | 0.000402 | 0.004598 | 0.000402 |
| 4 | 1 | 22.77 | 0.000940 | 0.004060 | 0.000940 |
| 5 | 1 | 47.77 | 0.001731 | 0.003269 | 0.001731 |
| 1 | 2 | 0.00 | 0.000000 | 0.005000 | 0.001000 |
| 2 | 2 | 4.15 | 0.000259 | 0.004741 | 0.001259 |
| 3 | 2 | 8.06 | 0.000437 | 0.004563 | 0.001437 |
| 4 | 2 | 22.77 | 0.001007 | 0.003993 | 0.002007 |
| 5 | 2 | 47.77 | 0.001792 | 0.003208 | 0.002792 |
| 1 | 3 | 0.00 | 0.000000 | 0.005000 | 0.002000 |
| 2 | 3 | 4.15 | 0.000262 | 0.004738 | 0.002262 |
| 3 | 3 | 8.06 | 0.000459 | 0.004541 | 0.002459 |
| 4 | 3 | 22.77 | 0.001049 | 0.003951 | 0.003049 |
| 5 | 3 | 47.77 | 0.001816 | 0.003184 | 0.003816 |
| end | | | | | |

## 3.2.2 Replication Assays Using Imidazole Buffer

Conditions: 5 mM tri-PNAs **A** and **B**, 0.2 M EDC, 0.4 M imidazole pH 7.5, 25 µl scale (see *Fig. 2* and *Table 2*).

$T = 0°$

| numb. | exp. | time[h] | t-t0 | a | t |
|---|---|---|---|---|---|
| 1 | 1 | 0.00 | 0.000000 | 0.00500 | 0.0000001 |
| 2 | 1 | 3.58 | 0.000097 | 0.00490 | 0.000097 |
| 3 | 1 | 6.58 | 0.000155 | 0.00485 | 0.000155 |
| 4 | 1 | 20.46 | 0.000385 | 0.00461 | 0.000385 |
| 5 | 1 | 29.77 | 0.000527 | 0.00447 | 0.000527 |
| 1 | 2 | 0.00 | 0.000000 | 0.00500 | 0.001500 |
| 2 | 2 | 3.58 | 0.000194 | 0.00481 | 0.001694 |
| 3 | 2 | 6.58 | 0.000227 | 0.00477 | 0.001727 |
| 4 | 2 | 20.46 | 0.000526 | 0.00447 | 0.002026 |
| 5 | 2 | 29.77 | 0.000673 | 0.00433 | 0.002173 |
| 1 | 3 | 0.00 | 0.000000 | 0.00500 | 0.002500 |
| 2 | 3 | 3.58 | 0.000182 | 0.00482 | 0.002682 |
| 3 | 3 | 6.58 | 0.000249 | 0.00475 | 0.002749 |
| 4 | 3 | 20.46 | 0.000509 | 0.00449 | 0.003009 |
| 5 | 3 | 29.77 | 0.000632 | 0.00437 | 0.003132 |
| end | | | | | |

*T* = 10°

| numb. | exp. | time[h] | t-t0 | a | t |
|---|---|---|---|---|---|
| 1 | 1 | 0.00 | 0.000000 | 0.005000 | 0.0000001 |
| 2 | 1 | 2.33 | 0.000041 | 0.004959 | 0.000041 |
| 3 | 1 | 6.63 | 0.000198 | 0.004802 | 0.000198 |
| 4 | 1 | 20.52 | 0.000569 | 0.004431 | 0.000569 |
| 5 | 1 | 29.80 | 0.000723 | 0.004277 | 0.000723 |
| 6 | 1 | 44.65 | 0.000922 | 0.004078 | 0.000922 |
| 1 | 2 | 0.00 | 0.000000 | 0.005000 | 0.001500 |
| 2 | 2 | 2.33 | 0.000237 | 0.004763 | 0.001737 |
| 3 | 2 | 6.63 | 0.000454 | 0.004546 | 0.001954 |
| 4 | 2 | 20.52 | 0.000856 | 0.004144 | 0.002356 |
| 5 | 2 | 29.80 | 0.001032 | 0.003968 | 0.002532 |
| 6 | 2 | 44.65 | 0.001176 | 0.003824 | 0.002676 |
| 1 | 3 | 0.00 | 0.000000 | 0.005000 | 0.002500 |
| 2 | 3 | 2.33 | 0.000223 | 0.004777 | 0.002723 |
| 3 | 3 | 6.63 | 0.000401 | 0.004599 | 0.002901 |
| 4 | 3 | 20.52 | 0.000871 | 0.004129 | 0.003371 |
| 5 | 3 | 29.80 | 0.001085 | 0.003915 | 0.003585 |
| 6 | 3 | 44.65 | 0.001281 | 0.003719 | 0.003781 |
| end | | | | | |

### 3.2.3 Replication Assays Involving Different Nucleophilic Catalysts

Conditions: 5 mM tri-PNAs **A** and **B**, 0.2 M EDC, 0.2 M MOPS pH 7.2, 0.2 M NaCl, 6 μl scale, exp.1-2: *T* = 10°, exp.3: *T* = r.t. (see *Fig. 3* and *Table 3*).

a: no catalyst

| numb. | exp. | time[h] | t-t0 | a | t |
|---|---|---|---|---|---|
| 1 | 1 | 0.00 | 0.000000 | 0.005000 | 0.0000001 |
| 2 | 1 | 1.35 | 0.000065 | 0.004935 | 0.000065 |
| 3 | 1 | 2.77 | 0.000084 | 0.004916 | 0.000084 |
| 4 | 1 | 5.09 | 0.000145 | 0.004855 | 0.000145 |
| 5 | 1 | 8.00 | 0.000203 | 0.004797 | 0.000203 |
| 6 | 1 | 23.22 | 0.000476 | 0.004524 | 0.000476 |
| 7 | 1 | 31.72 | 0.000572 | 0.004428 | 0.000572 |
| 1 | 2 | 0.00 | 0.000000 | 0.005000 | 0.000500 |
| 2 | 2 | 1.35 | 0.000031 | 0.004969 | 0.000531 |
| 3 | 2 | 2.77 | 0.000099 | 0.004901 | 0.000599 |
| 4 | 2 | 5.09 | 0.000227 | 0.004773 | 0.000727 |
| 5 | 2 | 8.00 | 0.000308 | 0.004692 | 0.000808 |
| 6 | 2 | 23.22 | 0.000721 | 0.004279 | 0.001221 |
| 7 | 2 | 31.72 | 0.000866 | 0.004134 | 0.001366 |
| 1 | 3 | 0.00 | 0.000000 | 0.005000 | 0.0000001 |
| 2 | 3 | 1.35 | 0.000112 | 0.004888 | 0.000112 |
| 3 | 3 | 2.77 | 0.000199 | 0.004801 | 0.000199 |
| 4 | 3 | 5.09 | 0.000327 | 0.004673 | 0.000327 |
| 5 | 3 | 8.00 | 0.000431 | 0.004569 | 0.000431 |
| 6 | 3 | 23.22 | 0.000719 | 0.004281 | 0.000719 |
| 7 | 3 | 31.72 | 0.000754 | 0.004246 | 0.000754 |
| end | | | | | |

b: 0.1 M Imidazole

| numb. | exp. | time[h] | t-t0 | a | t |
|---|---|---|---|---|---|
| 1 | 1 | 0.00 | 0.000000 | 0.005000 | 0.0000001 |
| 2 | 1 | 2.18 | 0.000040 | 0.004960 | 0.000040 |
| 3 | 1 | 4.13 | 0.000077 | 0.004923 | 0.000077 |
| 4 | 1 | 6.13 | 0.000152 | 0.004848 | 0.000152 |
| 5 | 1 | 8.47 | 0.000199 | 0.004801 | 0.000199 |
| 6 | 1 | 23.22 | 0.000765 | 0.004235 | 0.000765 |
| 7 | 1 | 31.28 | 0.000954 | 0.004046 | 0.000954 |
| 1 | 2 | 0.00 | 0.000000 | 0.005000 | 0.000500 |
| 2 | 2 | 1.13 | 0.000003 | 0.004997 | 0.000503 |
| 3 | 2 | 2.18 | 0.000081 | 0.004919 | 0.000581 |
| 4 | 2 | 4.13 | 0.000177 | 0.004823 | 0.000677 |
| 5 | 2 | 6.13 | 0.000241 | 0.004759 | 0.000741 |
| 6 | 2 | 8.47 | 0.000384 | 0.004616 | 0.000884 |
| 7 | 2 | 23.22 | 0.000928 | 0.004072 | 0.001428 |
| 8 | 2 | 31.28 | 0.001052 | 0.003948 | 0.001552 |
| 1 | 3 | 0.00 | 0.000000 | 0.005000 | 0.0000001 |
| 2 | 3 | 1.13 | 0.000020 | 0.004980 | 0.000020 |
| 3 | 3 | 2.18 | 0.000062 | 0.004938 | 0.000062 |
| 4 | 3 | 4.13 | 0.000180 | 0.004820 | 0.000180 |
| 5 | 3 | 6.13 | 0.000304 | 0.004696 | 0.000304 |
| 6 | 3 | 8.47 | 0.000319 | 0.004681 | 0.000319 |
| 7 | 3 | 23.22 | 0.000687 | 0.004313 | 0.000687 |
| 8 | 3 | 31.28 | 0.000787 | 0.004213 | 0.000787 |
| end | | | | | |

c: 0.1 M 1-Methylimidazole

| numb. | exp. | time[h] | t-t0 | a | t |
|---|---|---|---|---|---|
| 1 | 1 | 0.00 | 0.000000 | 0.005000 | 0.0000001 |
| 2 | 1 | 1.11 | 0.000038 | 0.004962 | 0.000038 |
| 3 | 1 | 2.21 | 0.000093 | 0.004907 | 0.000093 |
| 4 | 1 | 4.05 | 0.000201 | 0.004799 | 0.000201 |
| 5 | 1 | 8.02 | 0.000404 | 0.004596 | 0.000404 |
| 6 | 1 | 24.35 | 0.000810 | 0.004190 | 0.000810 |
| 7 | 1 | 32.40 | 0.000975 | 0.004025 | 0.000975 |
| 1 | 2 | 0.00 | 0.000000 | 0.005000 | 0.000500 |
| 2 | 2 | 1.11 | 0.000005 | 0.004995 | 0.000505 |
| 3 | 2 | 2.21 | 0.000120 | 0.004880 | 0.000620 |
| 4 | 2 | 4.05 | 0.000182 | 0.004818 | 0.000682 |
| 5 | 2 | 8.02 | 0.000314 | 0.004686 | 0.000814 |
| 6 | 2 | 24.35 | 0.000738 | 0.004262 | 0.001238 |
| 7 | 2 | 32.40 | 0.000919 | 0.004081 | 0.001419 |
| 1 | 3 | 0.00 | 0.000000 | 0.005000 | 0.0000001 |
| 2 | 3 | 1.11 | 0.000335 | 0.004665 | 0.000335 |
| 3 | 3 | 2.21 | 0.000485 | 0.004515 | 0.000485 |
| 4 | 3 | 4.05 | 0.000975 | 0.004025 | 0.000975 |
| 5 | 3 | 8.02 | 0.001651 | 0.003349 | 0.001651 |
| 6 | 3 | 24.35 | 0.002905 | 0.002095 | 0.002905 |
| 7 | 3 | 32.40 | 0.003270 | 0.001730 | 0.003270 |
| end | | | | | |

d: 0.1 M Pyridine

| numb. | exp. | time[h] | t-t0 | a | t |
|---|---|---|---|---|---|
| 1 | 1 | 0.00 | 0.000000 | 0.005000 | 0.0000001 |
| 2 | 1 | 1.02 | 0.000062 | 0.004938 | 0.000062 |
| 3 | 1 | 2.10 | 0.000056 | 0.004944 | 0.000056 |
| 4 | 1 | 4.15 | 0.000156 | 0.004844 | 0.000156 |
| 5 | 1 | 7.72 | 0.000270 | 0.004730 | 0.000270 |
| 6 | 1 | 22.37 | 0.000820 | 0.004180 | 0.000820 |
| 1 | 2 | 0.00 | 0.000000 | 0.005000 | 0.000500 |
| 2 | 2 | 1.02 | 0.000078 | 0.004922 | 0.000578 |
| 3 | 2 | 2.10 | 0.000102 | 0.004898 | 0.000602 |
| 4 | 2 | 4.15 | 0.000194 | 0.004806 | 0.000694 |
| 5 | 2 | 7.72 | 0.000356 | 0.004644 | 0.000856 |
| 6 | 2 | 22.37 | 0.000900 | 0.004100 | 0.001400 |
| 1 | 3 | 0.00 | 0.000000 | 0.005000 | 0.0000001 |
| 2 | 3 | 1.02 | 0.000343 | 0.004657 | 0.000343 |
| 3 | 3 | 2.10 | 0.000669 | 0.004331 | 0.000669 |
| 4 | 3 | 4.15 | 0.001184 | 0.003816 | 0.001184 |
| 5 | 3 | 7.72 | 0.001882 | 0.003118 | 0.001882 |
| 6 | 3 | 22.37 | 0.002968 | 0.002032 | 0.002968 |
| end | | | | | |

e: 0.1 M HOAt

| numb. | exp. | time[h] | t-t0 | a | t |
|---|---|---|---|---|---|
| 1 | 1 | 0.00 | 0.000000 | 0.005000 | 0.0000001 |
| 2 | 1 | 0.56 | 0.001652 | 0.003348 | 0.001652 |
| 3 | 1 | 1.11 | 0.001996 | 0.003004 | 0.001996 |
| 4 | 1 | 2.19 | 0.002671 | 0.002329 | 0.002671 |
| 5 | 1 | 4.24 | 0.003300 | 0.001700 | 0.003300 |
| 6 | 1 | 7.81 | 0.003805 | 0.001195 | 0.003805 |
| 7 | 1 | 22.46 | 0.004443 | 0.000557 | 0.004443 |
| 1 | 2 | 0.00 | 0.000000 | 0.005000 | 0.000500 |
| 2 | 2 | 0.56 | 0.001567 | 0.003433 | 0.002067 |
| 3 | 2 | 1.11 | 0.001881 | 0.003119 | 0.002381 |
| 4 | 2 | 2.19 | 0.002560 | 0.002440 | 0.003060 |
| 5 | 2 | 4.24 | 0.003358 | 0.001642 | 0.003858 |
| 6 | 2 | 7.81 | 0.003538 | 0.001462 | 0.004038 |
| 7 | 2 | 22.46 | 0.004047 | 0.000953 | 0.004547 |
| 1 | 3 | 0.00 | 0.000000 | 0.005000 | 0.0000001 |
| 2 | 3 | 0.56 | 0.002005 | 0.002995 | 0.002005 |
| 3 | 3 | 1.11 | 0.002470 | 0.002530 | 0.002470 |
| 4 | 3 | 2.19 | 0.002975 | 0.002025 | 0.002975 |
| 5 | 3 | 4.24 | 0.003050 | 0.001950 | 0.003050 |
| 6 | 3 | 7.81 | 0.003115 | 0.001885 | 0.003115 |
| 7 | 3 | 22.46 | 0.003404 | 0.001596 | 0.003404 |
| end | | | | | |

*3.2.5 Replication Assays Involving Different Salts*

Conditions: 5 mM tri-PNAs **A** and **B**, 0.2 M EDC, 0.2 M MOPS pH 7.6, 0.1 M imidazole, 6 µl scale, exp.1-2: *T* = 10°, exp.3: *T* = r.t. (see *Fig. 4* and *Table 4*).

a: No Additional Salt

| numb. | exp. | time[h] | t-t0 | a | t |
|---|---|---|---|---|---|
| 1 | 1 | 0.00 | 0.000000 | 0.005000 | 0.0000001 |
| 2 | 1 | 1.02 | 0.000034 | 0.004966 | 0.000034 |
| 3 | 1 | 2.15 | 0.000034 | 0.004966 | 0.000034 |
| 4 | 1 | 4.17 | 0.000091 | 0.004909 | 0.000091 |
| 5 | 1 | 7.70 | 0.000177 | 0.004823 | 0.000177 |
| 6 | 1 | 21.07 | 0.000494 | 0.004506 | 0.000494 |
| 7 | 1 | 28.82 | 0.000701 | 0.004299 | 0.000701 |
| 1 | 2 | 0.00 | 0.000000 | 0.005000 | 0.000500 |
| 2 | 2 | 1.02 | -0.00001 | 0.005011 | 0.000489 |
| 3 | 2 | 2.15 | 0.000067 | 0.004933 | 0.000567 |
| 4 | 2 | 4.17 | 0.000166 | 0.004834 | 0.000666 |
| 5 | 2 | 7.70 | 0.000289 | 0.004711 | 0.000789 |
| 6 | 2 | 21.07 | 0.000718 | 0.004282 | 0.001218 |
| 7 | 2 | 28.82 | 0.000901 | 0.004099 | 0.001401 |
| 1 | 3 | 0.00 | 0.000000 | 0.005000 | 0.0000001 |
| 2 | 3 | 1.02 | 0.000042 | 0.004958 | 0.000042 |
| 3 | 3 | 2.15 | 0.000115 | 0.004885 | 0.000115 |
| 4 | 3 | 4.17 | 0.000127 | 0.004873 | 0.000127 |
| 5 | 3 | 7.70 | 0.000293 | 0.004707 | 0.000293 |
| 6 | 3 | 21.07 | 0.000405 | 0.004595 | 0.000405 |
| 7 | 3 | 28.82 | 0.000456 | 0.004544 | 0.000456 |
| end | | | | | |

b: 0.2 M NaCl

| numb. | exp. | time[h] | t-t0 | a | t |
|---|---|---|---|---|---|
| 1 | 1 | 0.00 | 0.000000 | 0.005000 | 0.0000001 |
| 2 | 1 | 1.07 | -0.000006 | 0.005006 | -0.000006 |
| 3 | 1 | 2.21 | 0.000021 | 0.004979 | 0.000021 |
| 4 | 1 | 4.22 | 0.000089 | 0.004911 | 0.000089 |
| 5 | 1 | 7.71 | 0.000221 | 0.004779 | 0.000221 |
| 6 | 1 | 21.19 | 0.000573 | 0.004427 | 0.000573 |
| 7 | 1 | 28.87 | 0.000686 | 0.004314 | 0.000686 |
| 1 | 2 | 0.00 | 0.000000 | 0.005000 | 0.000500 |
| 2 | 2 | 1.07 | 0.000040 | 0.004960 | 0.000540 |
| 3 | 2 | 2.21 | 0.000086 | 0.004914 | 0.000586 |
| 4 | 2 | 4.22 | 0.000221 | 0.004779 | 0.000721 |
| 5 | 2 | 7.71 | 0.000393 | 0.004607 | 0.000893 |
| 6 | 2 | 21.19 | 0.000731 | 0.004269 | 0.001231 |
| 7 | 2 | 28.87 | 0.000936 | 0.004064 | 0.001436 |
| 1 | 3 | 0.00 | 0.000000 | 0.005000 | 0.0000001 |
| 2 | 3 | 1.07 | 0.000016 | 0.004984 | 0.000016 |
| 3 | 3 | 2.21 | 0.000034 | 0.004966 | 0.000034 |
| 4 | 3 | 4.22 | 0.000098 | 0.004902 | 0.000098 |
| 5 | 3 | 7.71 | 0.000241 | 0.004759 | 0.000241 |
| 6 | 3 | 21.19 | 0.000514 | 0.004486 | 0.000514 |
| 7 | 3 | 28.87 | 0.000566 | 0.004434 | 0.000566 |
| end | | | | | |

c: 0.2 M Na$_2$SO$_4$

| numb. | exp. | time[h] | t-t0 | a | t |
|---|---|---|---|---|---|
| 1 | 1 | 0.00 | 0.000000 | 0.005000 | 0.0000001 |
| 2 | 1 | 0.94 | -0.000003 | 0.005003 | -0.000003 |
| 3 | 1 | 1.92 | -0.000019 | 0.005019 | -0.000019 |
| 4 | 1 | 3.97 | 0.000046 | 0.004954 | 0.000046 |
| 5 | 1 | 6.64 | 0.000101 | 0.004899 | 0.000101 |
| 6 | 1 | 21.89 | 0.000388 | 0.004612 | 0.000388 |
| 7 | 1 | 29.92 | 0.000536 | 0.004464 | 0.000536 |
| 1 | 2 | 0.00 | 0.000000 | 0.005000 | 0.000500 |
| 2 | 2 | 0.94 | 0.000107 | 0.004893 | 0.000607 |
| 3 | 2 | 1.92 | 0.000083 | 0.004917 | 0.000583 |
| 4 | 2 | 3.97 | 0.000154 | 0.004846 | 0.000654 |
| 5 | 2 | 6.64 | 0.000190 | 0.004810 | 0.000690 |
| 6 | 2 | 21.89 | 0.000600 | 0.004400 | 0.001100 |
| 7 | 2 | 29.92 | 0.000778 | 0.004222 | 0.001278 |
| 1 | 3 | 0.00 | 0.000000 | 0.005000 | 0.0000001 |
| 2 | 3 | 0.94 | 0.000005 | 0.004995 | 0.000005 |
| 3 | 3 | 1.92 | 0.000046 | 0.004954 | 0.000046 |
| 4 | 3 | 3.97 | 0.000071 | 0.004929 | 0.000071 |
| 5 | 3 | 6.64 | 0.000091 | 0.004909 | 0.000091 |
| 6 | 3 | 21.89 | 0.000245 | 0.004755 | 0.000245 |
| 7 | 3 | 29.92 | 0.000268 | 0.004732 | 0.000268 |
| end | | | | | |

d: 0.2 M NaI

| numb. | exp. | time[h] | t-t0 | a | t |
|---|---|---|---|---|---|
| 1 | 1 | 0.00 | 0.000000 | 0.005000 | 0.0000001 |
| 2 | 1 | 0.97 | 0.000018 | 0.004982 | 0.000018 |
| 3 | 1 | 1.96 | -0.000010 | 0.005010 | -0.000010 |
| 4 | 1 | 4.01 | 0.000084 | 0.004916 | 0.000084 |
| 5 | 1 | 6.67 | 0.000172 | 0.004828 | 0.000172 |
| 6 | 1 | 21.89 | 0.000677 | 0.004323 | 0.000677 |
| 7 | 1 | 29.92 | 0.000905 | 0.004095 | 0.000905 |
| 1 | 2 | 0.00 | 0.000000 | 0.005000 | 0.000500 |
| 2 | 2 | 0.97 | 0.000034 | 0.004966 | 0.000534 |
| 3 | 2 | 1.96 | 0.000074 | 0.004926 | 0.000574 |
| 4 | 2 | 4.01 | 0.000210 | 0.004790 | 0.000710 |
| 5 | 2 | 6.67 | 0.000326 | 0.004674 | 0.000826 |
| 6 | 2 | 21.89 | 0.000948 | 0.004052 | 0.001448 |
| 7 | 2 | 29.92 | 0.001159 | 0.003841 | 0.001659 |
| 1 | 3 | 0.00 | 0.000000 | 0.005000 | 0.0000001 |
| 2 | 3 | 0.97 | 0.000021 | 0.004979 | 0.000021 |
| 3 | 3 | 1.96 | 0.000067 | 0.004933 | 0.000067 |
| 4 | 3 | 4.01 | 0.000152 | 0.004848 | 0.000152 |
| 5 | 3 | 6.67 | 0.000335 | 0.004665 | 0.000335 |
| 6 | 3 | 21.89 | 0.000465 | 0.004535 | 0.000465 |
| end | | | | | |

*3.2.6 Replication Assay Involving 0.4 M EDC*

Conditions: 5 mM tri-PNAs **A** and **B**, 0.4 M EDC, 0.2 M MOPS pH 7.6, 0.2 M NaCl, 6 µl scale, exp.1-2: *T* = 10°, exp.3: *T* = r.t. (see *Fig. 5* and *Table 5*).

| numb. | exp. | time[h] | t-t0 | a | t |
|---|---|---|---|---|---|
| 1 | 1 | 0.00 | 0.000000 | 0.005000 | 0.0000001 |
| 2 | 1 | 1.12 | 0.000080 | 0.004920 | 0.000080 |
| 3 | 1 | 2.07 | 0.000108 | 0.004892 | 0.000108 |
| 4 | 1 | 4.14 | 0.000224 | 0.004776 | 0.000224 |
| 5 | 1 | 8.49 | 0.000574 | 0.004426 | 0.000574 |
| 6 | 1 | 22.27 | 0.001209 | 0.003791 | 0.001209 |
| 1 | 2 | 0.00 | 0.000000 | 0.005000 | 0.000500 |
| 2 | 2 | 1.12 | 0.000118 | 0.004882 | 0.000618 |
| 3 | 2 | 2.07 | 0.000226 | 0.004774 | 0.000726 |
| 4 | 2 | 4.14 | 0.000412 | 0.004588 | 0.000912 |
| 5 | 2 | 8.49 | 0.000746 | 0.004254 | 0.001246 |
| 6 | 2 | 22.27 | 0.001444 | 0.003556 | 0.001944 |
| 1 | 3 | 0.00 | 0.000000 | 0.005000 | 0.0000001 |
| 2 | 3 | 1.12 | 0.000101 | 0.004899 | 0.000101 |
| 3 | 3 | 2.07 | 0.000197 | 0.004803 | 0.000197 |
| 4 | 3 | 4.14 | 0.000290 | 0.004710 | 0.000290 |
| 5 | 3 | 8.49 | 0.000515 | 0.004485 | 0.000515 |
| 6 | 3 | 22.27 | 0.000828 | 0.004172 | 0.000828 |
| end | | | | | |

comparable experiment at 0.2 M EDC

| numb. | exp. | time[h] | t-t0 | a | t |
|---|---|---|---|---|---|
| 1 | 1 | 0.00 | 0.000000 | 0.005000 | 0.0000001 |
| 2 | 1 | 1.07 | -0.000006 | 0.005006 | -0.000006 |
| 3 | 1 | 2.21 | 0.000021 | 0.004979 | 0.000021 |
| 4 | 1 | 4.22 | 0.000089 | 0.004911 | 0.000089 |
| 5 | 1 | 7.71 | 0.000221 | 0.004779 | 0.000221 |
| 6 | 1 | 21.19 | 0.000573 | 0.004427 | 0.000573 |
| 1 | 2 | 0.00 | 0.000000 | 0.005000 | 0.000500 |
| 2 | 2 | 1.07 | 0.000040 | 0.004960 | 0.000540 |
| 3 | 2 | 2.21 | 0.000086 | 0.004914 | 0.000586 |
| 4 | 2 | 4.22 | 0.000221 | 0.004779 | 0.000721 |
| 5 | 2 | 7.71 | 0.000393 | 0.004607 | 0.000893 |
| 6 | 2 | 21.19 | 0.000731 | 0.004269 | 0.001231 |
| 1 | 3 | 0.00 | 0.000000 | 0.005000 | 0.0000001 |
| 2 | 3 | 1.07 | 0.000016 | 0.004984 | 0.000016 |
| 3 | 3 | 2.21 | 0.000034 | 0.004966 | 0.000034 |
| 4 | 3 | 4.22 | 0.000098 | 0.004902 | 0.000098 |
| 5 | 3 | 7.71 | 0.000241 | 0.004759 | 0.000241 |
| 6 | 3 | 21.19 | 0.000514 | 0.004486 | 0.000514 |
| end | | | | | |

*3.2.7 Replication Assay at pH 6.6*

Conditions: 5 mM tri-PNAs **A** and **B**, 0.2 M EDC, 0.2 M MOPS, 0.2 M NaCl, 0.1 M imidazole, 6 μl scale, exp.1-2: *T* = 10°, exp.3: *T* = r.t. (see *Fig. 6* and *Table 6*).

| numb. | exp. | time[h] | t-t0 | a | t |
|---|---|---|---|---|---|
| 1 | 1 | 0.00 | 0.000000 | 0.005000 | 0.0000001 |
| 2 | 1 | 1.02 | 0.000046 | 0.004954 | 0.000046 |
| 3 | 1 | 1.98 | 0.000131 | 0.004869 | 0.000131 |
| 4 | 1 | 4.05 | 0.000312 | 0.004688 | 0.000312 |
| 5 | 1 | 8.38 | 0.000916 | 0.004084 | 0.000916 |
| 1 | 2 | 0.00 | 0.000000 | 0.005000 | 0.000500 |
| 2 | 2 | 1.02 | 0.000012 | 0.004988 | 0.000512 |
| 3 | 2 | 1.98 | 0.000163 | 0.004837 | 0.000663 |
| 4 | 2 | 4.05 | 0.000546 | 0.004454 | 0.001046 |
| 5 | 2 | 8.38 | 0.001153 | 0.003847 | 0.001653 |
| 1 | 3 | 0.00 | 0.000000 | 0.005000 | 0.0000001 |
| 2 | 3 | 1.02 | 0.000105 | 0.004895 | 0.000105 |
| 3 | 3 | 1.98 | 0.000342 | 0.004658 | 0.000342 |
| 4 | 3 | 4.05 | 0.000583 | 0.004417 | 0.000583 |
| 5 | 3 | 8.38 | 0.001168 | 0.003832 | 0.001168 |
| end | | | | | |

comparable experiment at pH 7.6: see 3.2.5 b

comparable experiment at pH 7.2: see 3.2.3 b

*3.2.8 Replication Assay Optimized for Fast Replication and High Conversion*

Conditions: 5 mM tri-PNAs **A** and **B**, 0.4 M EDC, 0.2 M MOPS pH 6.6, 0.1 M imidazole, 10°, 6 µl scale (see *Fig. 7* and *Table 7*).

| numb. | exp. | time[h] | t-t0 | a | t |
|---|---|---|---|---|---|
| 1 | 1 | 0.00 | 0.000000 | 0.005000 | 0.0000001 |
| 2 | 1 | 1.09 | 0.000185 | 0.004815 | 0.000185 |
| 3 | 1 | 2.16 | 0.000378 | 0.004622 | 0.000378 |
| 4 | 1 | 3.52 | 0.000730 | 0.004270 | 0.000730 |
| 5 | 1 | 5.16 | 0.001171 | 0.003829 | 0.001171 |
| 6 | 1 | 7.05 | 0.001647 | 0.003353 | 0.001647 |
| 7 | 1 | 23.47 | 0.003277 | 0.001723 | 0.003277 |
| 1 | 2 | 0.00 | 0.000000 | 0.005000 | 0.000250 |
| 2 | 2 | 1.09 | 0.000246 | 0.004754 | 0.000496 |
| 3 | 2 | 2.16 | 0.000537 | 0.004463 | 0.000787 |
| 4 | 2 | 3.52 | 0.000895 | 0.004105 | 0.001145 |
| 5 | 2 | 5.16 | 0.001421 | 0.003579 | 0.001671 |
| 6 | 2 | 7.05 | 0.001851 | 0.003149 | 0.002101 |
| 7 | 2 | 23.47 | 0.003310 | 0.001690 | 0.003560 |
| 1 | 3 | 0.00 | 0.000000 | 0.005000 | 0.000500 |
| 2 | 3 | 1.09 | 0.000255 | 0.004745 | 0.000755 |
| 3 | 3 | 2.16 | 0.000593 | 0.004407 | 0.001093 |
| 4 | 3 | 3.52 | 0.000965 | 0.004035 | 0.001465 |
| 5 | 3 | 5.16 | 0.001450 | 0.003550 | 0.001950 |
| 6 | 3 | 7.05 | 0.001820 | 0.003180 | 0.002320 |
| 7 | 3 | 23.47 | 0.003498 | 0.001502 | 0.003998 |
| 1 | 4 | 0.00 | 0.000000 | 0.005000 | 0.001000 |
| 2 | 4 | 1.09 | 0.000277 | 0.004723 | 0.001277 |
| 3 | 4 | 2.16 | 0.000614 | 0.004386 | 0.001614 |
| 4 | 4 | 3.52 | 0.001033 | 0.003967 | 0.002033 |
| 5 | 4 | 5.16 | 0.001603 | 0.003397 | 0.002603 |
| 6 | 4 | 7.05 | 0.001951 | 0.003049 | 0.002951 |
| 7 | 4 | 23.47 | 0.003441 | 0.001559 | 0.004441 |
| end | | | | | |

*3.2.9 Replication Assays Involving Different Imidazole Concentrations*

Conditions: 5 mM tri-PNAs **A** and **B**, 0.2 M EDC, 0.2 M MOPS pH 7.2, 0.2 M NaI, 10°, 6 µl scale (see *Fig. 8* and *Table 8*).

a: 0.1 M Imidazol

| numb. | exp. | time[h] | t-t0 | a | t |
|---|---|---|---|---|---|
| 1 | 1 | 0.00 | 0.000000 | 0.005000 | 0.0000001 |
| 2 | 1 | 0.87 | 0.000092 | 0.004908 | 0.000092 |
| 3 | 1 | 1.50 | 0.000106 | 0.004894 | 0.000106 |
| 4 | 1 | 3.17 | 0.000288 | 0.004712 | 0.000288 |
| 5 | 1 | 5.67 | 0.000515 | 0.004485 | 0.000515 |
| 6 | 1 | 9.12 | 0.000829 | 0.004171 | 0.000829 |
| 7 | 1 | 23.63 | 0.001792 | 0.003208 | 0.001792 |
| 1 | 2 | 0.00 | 0.000000 | 0.005000 | 0.000500 |
| 2 | 2 | 0.87 | 0.000070 | 0.004930 | 0.000570 |
| 3 | 2 | 1.50 | 0.000139 | 0.004861 | 0.000639 |
| 4 | 2 | 3.17 | 0.000361 | 0.004639 | 0.000861 |
| 5 | 2 | 5.67 | 0.000683 | 0.004317 | 0.001183 |
| 6 | 2 | 9.12 | 0.001012 | 0.003988 | 0.001512 |
| 7 | 2 | 23.63 | 0.002050 | 0.002950 | 0.002550 |
| end | | | | | |

b: 0.2 M Imidazole

| numb. | exp. | time[h] | t-t0 | a | t |
|---|---|---|---|---|---|
| 1 | 1 | 0.00 | 0.000000 | 0.005000 | 0.0000001 |
| 2 | 1 | 0.75 | 0.000083 | 0.004917 | 0.000083 |
| 3 | 1 | 1.38 | 0.000055 | 0.004945 | 0.000055 |
| 4 | 1 | 3.05 | 0.000202 | 0.004798 | 0.000202 |
| 5 | 1 | 5.55 | 0.000383 | 0.004617 | 0.000383 |
| 6 | 1 | 9.00 | 0.000484 | 0.004516 | 0.000484 |
| 7 | 1 | 23.52 | 0.001119 | 0.003881 | 0.001119 |
| 1 | 2 | 0.00 | 0.000000 | 0.005000 | 0.000500 |
| 2 | 2 | 0.75 | 0.000075 | 0.004925 | 0.000575 |
| 3 | 2 | 3.05 | 0.000258 | 0.004742 | 0.000758 |
| 4 | 2 | 5.55 | 0.000518 | 0.004482 | 0.001018 |
| 5 | 2 | 9.00 | 0.000734 | 0.004266 | 0.001234 |
| 6 | 2 | 23.52 | 0.001301 | 0.003699 | 0.001801 |
| end | | | | | |

### 3.2.10 Replication Assay in the Eutectic Phase System at -19°

Conditions: 5 mM tri-PNAs **A** and **B**, 0.2 M EDC, 0.2 M MOPS pH 7.6, 0.2 M NaCl, 0.1 M imidazole, -19°, 12 µl scale (see *Fig. 9* and *Table 9*).

| numb. | exp. | time[h] | t-t0 | a | t |
|---|---|---|---|---|---|
| 1 | 1 | 0.00 | 0.000000 | 0.005000 | 0.0000001 |
| 2 | 1 | 2.10 | -0.000002 | 0.005002 | -0.000002 |
| 3 | 1 | 4.32 | 0.000050 | 0.004950 | 0.000050 |
| 4 | 1 | 7.20 | 0.000092 | 0.004908 | 0.000092 |
| 5 | 1 | 23.45 | 0.000392 | 0.004608 | 0.000392 |
| 6 | 1 | 30.87 | 0.000494 | 0.004506 | 0.000494 |
| 7 | 1 | 48.70 | 0.000763 | 0.004237 | 0.000763 |
| 1 | 2 | 0.00 | 0.000000 | 0.005000 | 0.000500 |
| 2 | 2 | 2.10 | 0.000057 | 0.004943 | 0.000557 |
| 3 | 2 | 4.32 | 0.000147 | 0.004853 | 0.000647 |
| 4 | 2 | 7.20 | 0.000177 | 0.004823 | 0.000677 |
| 5 | 2 | 30.87 | 0.000542 | 0.004458 | 0.001042 |
| 6 | 2 | 48.70 | 0.000831 | 0.004169 | 0.001331 |
| end | | | | | |

comparable experiment at 10°: see 3.2.6

*3.2.11 Replication Assays Involving PEG*

Conditions: 5 mM tri-PNAs **A** and **B**, 0.2 M EDC, 0.2 M MOPS pH 7.6, 0.2 M NaCl, 0.1 M imidazole, 20 % PEG, 6 μl scale, exp.1-2: $T$ = 10°, exp.3: $T$ = r.t. (see *Fig. 10* and *Table 10*).

a: PEG3350

| numb. | exp. | time[h] | t-t0 | a | t |
|---|---|---|---|---|---|
| 1 | 1 | 0.00 | 0.000000 | 0.005000 | 0.0000001 |
| 2 | 1 | 1.03 | -0.000023 | 0.005023 | -0.000023 |
| 3 | 1 | 2.02 | 0.000023 | 0.004977 | 0.000023 |
| 4 | 1 | 4.07 | 0.000128 | 0.004872 | 0.000128 |
| 5 | 1 | 7.88 | 0.000340 | 0.004660 | 0.000340 |
| 6 | 1 | 21.53 | 0.001081 | 0.003919 | 0.001081 |
| 1 | 2 | 0.00 | 0.000000 | 0.005000 | 0.000500 |
| 2 | 2 | 1.03 | -0.000006 | 0.005006 | 0.000494 |
| 3 | 2 | 2.02 | 0.000076 | 0.004924 | 0.000576 |
| 4 | 2 | 4.07 | 0.000283 | 0.004717 | 0.000783 |
| 5 | 2 | 7.88 | 0.000572 | 0.004428 | 0.001072 |
| 6 | 2 | 21.53 | 0.001403 | 0.003597 | 0.001903 |
| 1 | 3 | 0.00 | 0.000000 | 0.005000 | 0.0000001 |
| 2 | 3 | 1.03 | 0.000010 | 0.004990 | 0.000010 |
| 3 | 3 | 2.02 | 0.000089 | 0.004911 | 0.000089 |
| 4 | 3 | 4.07 | 0.000249 | 0.004751 | 0.000249 |
| 5 | 3 | 7.88 | 0.000402 | 0.004598 | 0.000402 |
| 6 | 3 | 21.53 | 0.000889 | 0.004111 | 0.000889 |
| end | | | | | |

b: PEG400

| numb. | exp. | time[h] | t-t0 | a | t |
|---|---|---|---|---|---|
| 1 | 1 | 0.00 | 0.000000 | 0.005000 | 0.0000001 |
| 2 | 1 | 1.15 | -0.000024 | 0.005024 | -0.000024 |
| 3 | 1 | 2.12 | 0.000007 | 0.004993 | 0.000007 |
| 4 | 1 | 4.17 | 0.000026 | 0.004974 | 0.000026 |
| 5 | 1 | 7.98 | 0.000091 | 0.004909 | 0.000091 |
| 6 | 1 | 21.65 | 0.000292 | 0.004708 | 0.000292 |
| 1 | 2 | 0.00 | 0.000000 | 0.005000 | 0.000700 |
| 2 | 2 | 1.15 | 0.000020 | 0.004980 | 0.000720 |
| 3 | 2 | 2.12 | 0.000052 | 0.004948 | 0.000752 |
| 4 | 2 | 4.17 | 0.000142 | 0.004858 | 0.000842 |
| 5 | 2 | 7.98 | 0.000294 | 0.004706 | 0.000994 |
| 6 | 2 | 21.65 | 0.000666 | 0.004334 | 0.001366 |
| 1 | 3 | 0.00 | 0.000000 | 0.005000 | 0.0000001 |
| 2 | 3 | 1.15 | -0.000016 | 0.005016 | -0.000016 |
| 3 | 3 | 2.12 | 0.000006 | 0.004994 | 0.000006 |
| 4 | 3 | 4.17 | 0.000070 | 0.004930 | 0.000070 |
| 5 | 3 | 7.98 | 0.000112 | 0.004888 | 0.000112 |
| 6 | 3 | 21.65 | 0.000257 | 0.004743 | 0.000257 |
| end | | | | | |

comparable experiment at 10°: see 3.2.6